\documentclass[11pt]{article}
\usepackage[margin=1in]{geometry}
\usepackage{graphicx}
\usepackage{float}
\usepackage{amsmath} 
\usepackage{amssymb}
\usepackage{amsfonts} 
\usepackage{amsthm}
\usepackage{algorithmic}
\usepackage{algorithm}
\usepackage[english]{babel}
\usepackage[latin1]{inputenc}
\usepackage{times}
\usepackage[T1]{fontenc}

\usepackage{amssymb}
\usepackage{amsmath}
\usepackage{amstext}
\usepackage{rotate}

\usepackage{amsfonts}
\usepackage{url}

\newcommand {\ignore}[1]{}               

\newcommand {\Id}{\mathbf{I}}

\newcommand {\br}[1]{\left(#1\right)}
\newcommand {\sqb}[1]{\left[#1\right]}
\newcommand {\cbr}[1]{\left\{#1 \right\}}
\newcommand {\nm}[1]{\Arrowvert\, #1 \,\Arrowvert}

\newcommand {\abs}[1]{\left\vert\, #1 \,\right\vert}

\newcommand {\ovr}[1]{\frac{1}{#1}}

\newcommand{\RR}{\mathbb{R}}    

\newcommand{\Acal}{\mathcal{A}} 

\newcommand{\argmin}{\mathop{\mathrm{argmin}\,}}
\newcommand{\argmax}{\mathop{\mathrm{argmax}\,}}

\newcommand{\eq}[1]{(\protect\ref{#1})}

\newcommand{\II}{\mbox{\boldmath $1$}}

\newtheorem{theorem}{Theorem}
\newtheorem{lemma}[theorem]{Lemma}
\graphicspath{{}}

\newcommand{\OMIT}[1]{}

\begin{document}

\title{The group fused Lasso for multiple change-point detection}
\author{Kevin Bleakley\\
INRIA Saclay, Orsay, France\\
\texttt{kevbleakley@gmail.com}
\and
Jean-Philippe Vert\\
Mines ParisTech CBIO, Fontainebleau, France\\
Institut Curie, Paris, France\\
INSERM U900, Paris, France\\
\texttt{Jean-Philippe.Vert@mines.org}
}
\date{}

\maketitle

\begin{abstract}
We present the group fused Lasso for detection of multiple change-points shared by a set of co-occurring one-dimensional signals. Change-points are detected by approximating the original signals with a constraint on the multidimensional total variation, leading to piecewise-constant approximations. Fast algorithms are proposed to solve the resulting optimization problems, either exactly or approximately. Conditions are given for consistency of both algorithms as the number of signals increases, and empirical evidence is provided to support the results on simulated and array comparative genomic hybridization data.

\end{abstract}

\section{Introduction}

Finding the place (or time) where most or all of a set of one-dimensional signals (or \emph{profiles}) jointly change in some specific way is an important question in several fields. A common situation is when we want to find change-points in a multidimensional signal, e.g., in audio and image processing \cite{Desobry2005online,Harchaoui2009regularized}, to detect intrusion in computer networks \cite{Tartakovsky2006novel,Levy-Leduc2009Detection}, or in financial and economics time series analysis \cite{Talih2005Structural}. Another important situation is when we are confronted with several 1-dimensional signals which we believe share common change-points, e.g., genomic profiles of a set of patients. The latter application is increasingly important in biology and medicine, in particular for the detection of copy-number variation along the genome \cite{Picard2005statistical}, or the analysis of microarray and genetic linkage studies \cite{Zhang2010Detecting}. The common thread in biological applications is the search for
data patterns shared by a set of individuals, such as cancer patients, at precise places on the genome; in particular, sudden changes in measured values. As opposed to the segmentation of multidimensional signals such as speech, where the dimension is fixed and collecting more data means having longer profiles, the length of signals in genomic studies (i.e., the number of probes measured along the genome) is fixed for a given technology while the number of signals (i.e., the number of individuals) can increase when we collect data about more patients. From a statistical point of view, it is therefore of interest to develop methods that identify multiple change-points shared by several signals that can benefit from increasing the number of signals.

There exists a vast literature on the change-point detection problem \cite{Basseville1993Detection,Brodsky1993Nonparametric}. Here we focus on computationally efficient methods to segment a multidimensional signal by approximating it with a piecewise-constant one, using quadratic error criteria. It is well-known that, in this case, the optimal segmentation of a $p$-dimensional signal of length $n$ into $k$ segments can be obtained in $O(n^2 p k)$ by dynamic programming \cite{Yao1988Estimating,Birge2001Gaussian,Lavielle2006Detection}. However, the quadratic complexity in $n$ is prohibitive in applications such as genomics, where $n$ can be in the order of $10^5$ to $10^7$ with current technology. An alternative to such \emph{global} procedures, which estimate change-points as solutions of a global optimization problem, are fast \emph{local} procedures such as binary segmentation \cite{Vostrikova1981Detection}, which detect breakpoints by iteratively applying a method for single change-point detection to the segments obtained after the previous change-point is detected. While such recursive methods can be extremely fast, in the order of  $O(np\log(k))$ when the single change-point detector is $O(np)$,  quality of segmentation is questionable when compared with global procedures \cite{Lavielle2005Adaptive}.

For $p=1$ (a single signal), an interesting alternative to these global and local procedures is to express the optimal segmentation as the solution of a convex optimization problem, using the (convex) total variation instead of the (non-convex) number of jumps to penalize a piecewise-constant function in order to approximate the original signal \cite{Rudin1992Nonlinear,Tibshirani2005Sparsity}. The resulting piecewise-constant approximation of the signal, defined as the global minimum of the objective function, benefits from theoretical guaranties in terms of correctly detecting change-points \cite{Harchaoui2008Catching,Rinaldo2009Properties}, and can be implemented efficiently in $O(nk)$ or $O(n\log(n))$ \cite{Friedman2007Pathwise,Harchaoui2008Catching,Hoefling2009path}.

In this paper we propose an extension of total-variation based methods for single signals to the multidimensional setting, in order to approximate a multidimensional signal with a piecewise-constant signal with multiple change-points. We define the approximation as the solution of a convex optimization problem which involves a quadratic approximation error penalized by the sum of the Euclidean norms of the multidimensional increments of the function. The problem can be reformulated as a group Lasso \cite{Yuan2006Model}, which we show how to solve exactly and efficiently. Alternatively, we provide an approximate yet often computationally faster solution to the problem using a group LARS procedure \cite{Yuan2006Model}. In the latter case, using the particular structure of the design matrix, we can find the first $k$ change-points in $O(npk)$, thus extending the method of \cite{Harchaoui2008Catching} to the multidimensional setting.

Unlike most previous theoretical investigations of change-point methods (e.g., \cite{Harchaoui2008Catching,Rinaldo2009Properties}), we are not interested in the case where the dimension $p$ is fixed and the length of the profiles $n$ increases, but in the opposite situation where $n$ is fixed and $p$ increases. Indeed, this corresponds to the case in genomics where, for example, $n$ would be the fixed number of probes used to measure a signal along the genome, and $p$ the number of samples or patients analyzed. We want to design a method that benefits from increasing $p$ in order to identify shared change-points, even though the signal-to-noise ratio may be very low within each signal. As a first step towards this question, we give conditions under which our method is able to consistently identify a single change-point as $p$ increases. We also show by simulation that the method is able to correctly identify multiple change-points as $p\rightarrow +\infty$, validating its relevance in practical settings. 

The paper is organized as follows. After fixing notation in Section \ref{sec:notations}, we present the group fused Lasso method in Section \ref{sec:formulation}. We propose two efficient algorithms to solve it in Section \ref{sec:implementation}, and discuss its theoretical properties in Section \ref{sec:theory}. Lastly, we provide an empirical evaluation of the method and a comparison with other methods in the study of copy number variations in cancer in Section \ref{sec:experiments}. A preliminary version of this paper was published in \cite{Vert2010Fast}.

\section{Notation}\label{sec:notations}
For any two integers $u\leq v$, we denote by $\sqb{u,v}$ the interval $\cbr{u,u+1,\ldots,v}$. For any $u\times v$ matrix $M$ we note $M_{i,j}$ its $(i,j)$-th entry, and $\|M\| = \sqrt{\sum_{i=1}^u\sum_{j=1}^v M_{i,j}^2}$ its Frobenius norm (or Euclidean norm in the case of vectors). For any subsets of indices $A = \br{a_1,\ldots,a_{|A|}}\in[1,u]^{|A|}$ and 
$B = \br{b_1,\ldots,b_{|B|}}\in[1,v]^{|B|}$, we denote by $M_{A,B}$ the $|A|\times|B|$ matrix with entries $M_{a_i,b_j}$ for $(i,j)\in[1,|A|]\times[1,|B|]$. For simplicity we will use $\bullet$ instead of $[1,u]$ or $[1,v]$, i.e., $A_{i,\bullet}$ is the $i$-th row of $A$ and $A_{\bullet, j}$ is the $j$-th column of $A$. We note $\II_{u,v}$ the $u \times v$ matrix of ones, and $\Id_p$ the $p\times p$ identity matrix.

\section{Formulation}\label{sec:formulation}
We consider $p$ real-valued profiles of length $n$, stored in an $n\times p$ matrix $Y$. The $i$-th profile $Y_{\bullet, i} = (Y_{1,i},\ldots,Y_{n,i})$ is the $i$-th column of $Y$. We model each profile as a piecewise-constant signal corrupted by noise, and assume that change-point locations tend to be shared across profiles. Our goal is to detect these shared change-points, and benefit from the possibly large number $p$ of profiles to increase the statistical power of change-point detection.

\subsection{Segmentation with a total variation penalty}

When $p=1$ (a single profile), a popular method to find change-points in a signal is to approximate it by a piecewise-constant function using a quadratic error criterion, i.e., to solve
\begin{equation}\label{eq:dp}
\min_{U \in \RR^{n}}  \nm{Y-U}^2 \quad \text{subject to} \quad\sum_{i=1}^{n-1} \delta(U_{i+1} - U_{i})\leq k\,,
\end{equation}
where $\delta$ is the Dirac function, equal to $0$ if its argument is null, $1$ otherwise. In other words, (\ref{eq:dp}) expresses the best approximation of $Y$ by a piecewise-constant profile $U$ with at most $k$ jumps. It is well-known that (\ref{eq:dp}) can be solved in $O(n^2 k)$ by dynamic programming \cite{Yao1988Estimating,Birge2001Gaussian,Lavielle2006Detection}. Although very fast when $n$ is of moderate size, the quadratic dependency in $n$ renders it impractical in current computers when $n$ reaches millions or more, which is often the case in many application such as segmentation of genomic profiles.

An alternative to the combinatorial optimization problem (\ref{eq:dp}) is to relax it to a convex optimization problem, by replacing the number of jumps by the convex total variation (TV) \cite{Rudin1992Nonlinear}, i.e., to consider:
\begin{equation}\label{eq:1Ddenoising}
\min_{U \in \RR^{n}} \frac{1}{2} \nm{Y-U}^2 + \lambda \sum_{i=1}^{n-1} \abs{U_{i+1} - U_{i}}\,.
\end{equation}
For a given $\lambda>0$, the solution $U\in\RR^n$ of (\ref{eq:1Ddenoising}) is again piecewise-constant. Recent work has shown that (\ref{eq:1Ddenoising}) can be solved much more efficiently than (\ref{eq:dp}): \cite{Friedman2007Pathwise} proposed a fast coordinate descent-like method, \cite{Harchaoui2008Catching} showed how to find the first $k$ change-points iteratively in $O(nk)$, and \cite{Hoefling2009path} proposed a $O(n\ln(n))$ method to find all change-points. Adding penalties proportional to the $\ell_1$ or $\ell_2$ norm of $U$ to (\ref{eq:1Ddenoising}) does not change the position of the change-points detected \cite{Tibshirani2005Sparsity,Mairal2010Online}, and the capacity of TV denoising to correctly identify 
change-points when $n$ increases has been investigated in \cite{Harchaoui2008Catching,Rinaldo2009Properties}.

Here, we propose to generalize TV denoising to multiple profiles by considering the following convex optimization problem, for $Y\in\RR^{n\times p}$:
\begin{equation}\label{eq:firstproblem}
\min_{U \in \RR^{n\times p}} \frac{1}{2}\nm{Y-U}^2 + \lambda \sum_{i=1}^{n-1} \nm{U_{i+1,\bullet} - U_{i,\bullet}}\,.
\end{equation}
The second term in (\ref{eq:firstproblem}) can be considered a multidimensional TV: it penalizes the sum of Euclidean norms of the increments of $U$, seen as a time-dependent multidimensional vector, and reduces to the classical 1-dimensional TV when $p=1$. Intuitively, when $\lambda$ increases, this penalty will enforce many increment vectors $U_{i+1,\bullet} - U_{i,\bullet}$ to collapse to $0$, just like the total variation in (\ref{eq:1Ddenoising}) in the case of $1$-dimensional signals. This implies that the positions of non-zero increments will be the same for all profiles. As a result, the solution to (\ref{eq:firstproblem}) provides an approximation of the profiles $Y$ by an $n\times p$ matrix of piecewise-constant profiles $U$ which share change-points. 

While (\ref{eq:firstproblem}) is a natural multidimensional generalization of the classical TV denoising method (\ref{eq:1Ddenoising}), we more generally investigate the following variant:
\begin{equation}\label{eq:weightedFGL}
\min_{U \in \RR^{n\times p}} \frac{1}{2}\nm{Y-U}^2 + \lambda \sum_{i=1}^{n-1} \frac{ \nm{U_{i+1,\bullet} - U_{i,\bullet}}}{d_i}\,,
\end{equation}
where $\br{d_i}_{i=1,\ldots,n-1}$ are position-dependant weights which affect the penalization of the jump differently at different positions. While (\ref{eq:weightedFGL}) boils down to (\ref{eq:firstproblem}) for uniform weights $d_i=1$, $i=1,\ldots,n-1$, we will see that the unweighted version suffers from boundary effects and that position-dependent schemes such as:
\begin{equation}\label{eq:weights}
\forall i\in[1,n-1],\quad d_i = \sqrt{\frac{n}{i(n-i)}}\,,
\end{equation}
are both theoretically and empirically better choices.

\begin{figure}[htfp]
\begin{center}
\includegraphics[width=5cm]{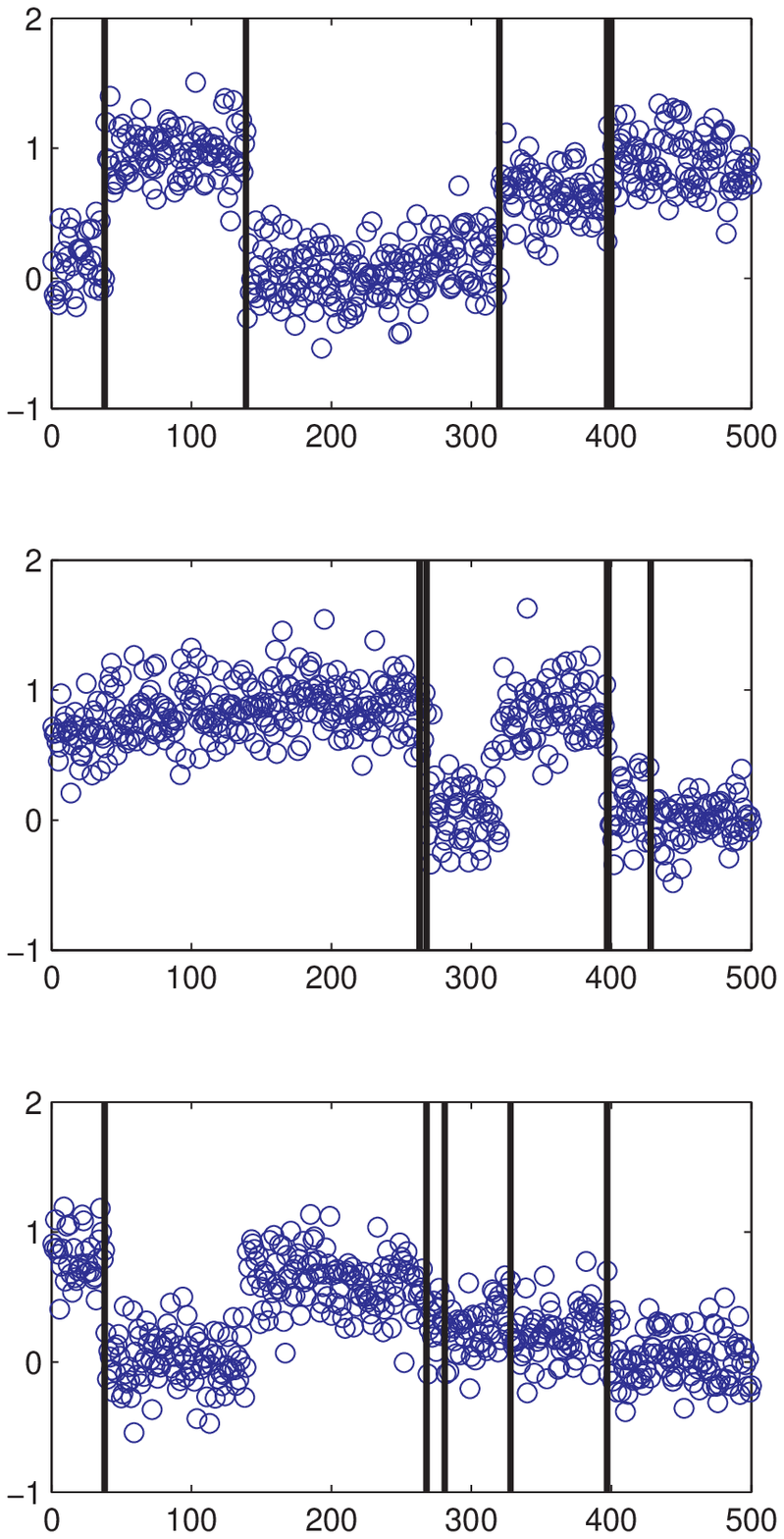}
\includegraphics[width=5cm]{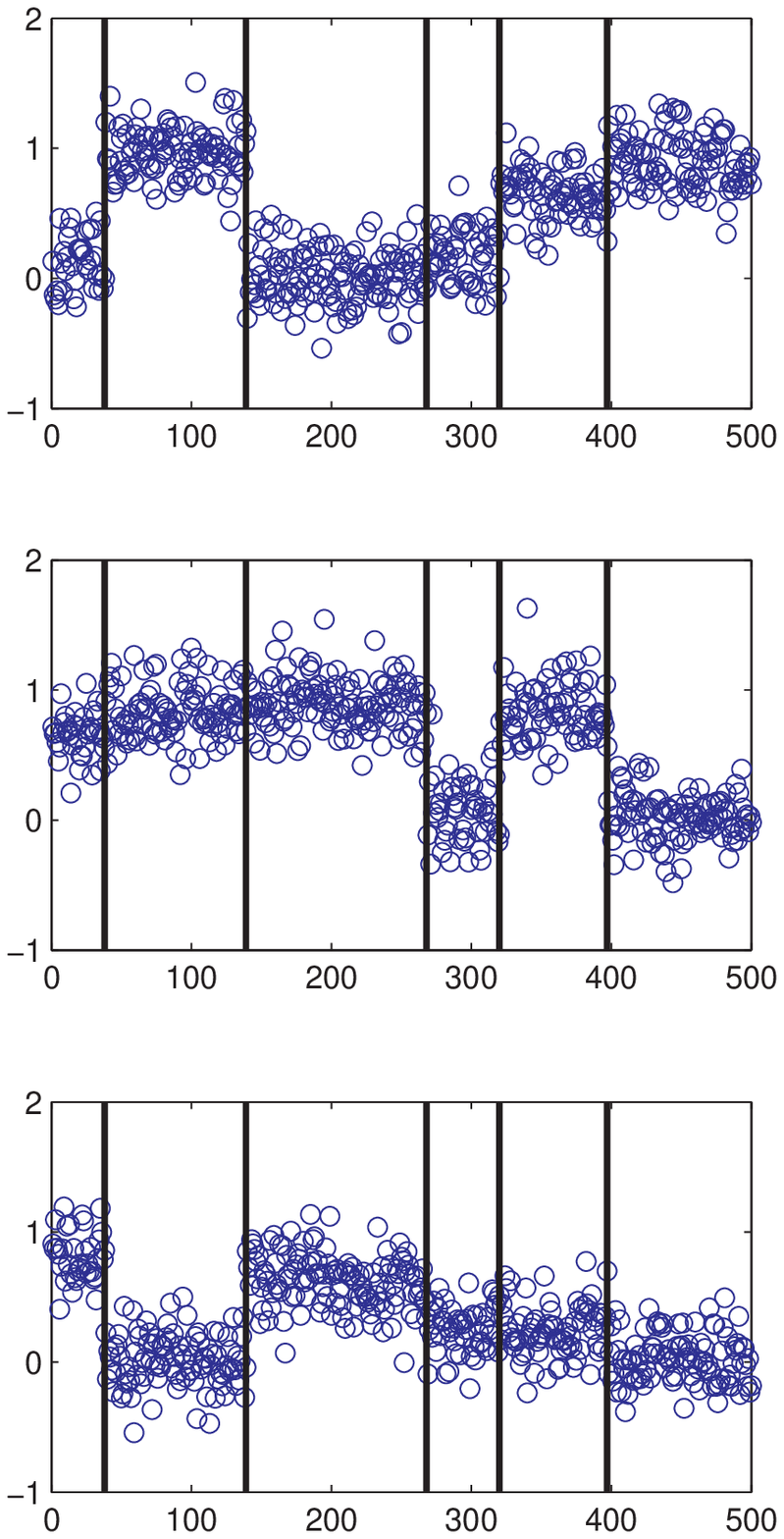}
\end{center}
\caption{First $5$ change-points detected on three simulated profiles by TV denoising of each profile (left) and by joint TV denoising (right).}
\label{fig:toy1}
\end{figure}

To illustrate the grouping effect of the penalty in (\ref{eq:weightedFGL}), Figure \ref{fig:toy1} compares the segmentation of three simulated profiles obtained with and without enforced sharing of change-points across profiles. We simulated three piecewise-constant signals corrupted by independent additive Gaussian noise. All profiles have length 500 and share the same $5$ change-points, though with different amplitudes, at positions 38, 139, 268, 320 and 397. On the left-hand side, we show the first $5$ change-points captured by TV denoising with weights (\ref{eq:weights}) applied to each signal independently. On the right, we show the first $5$ change-points captured by formulation (\ref{eq:weightedFGL}). We see that the latter formulation finds the correct change-points, whereas treating each profile independently leads to errors. For example, the first two change-points have a small amplitude in the second profile and are therefore very difficult to detect from the profile only, while they are very apparent in the first and third profiles.

\subsection{Reformulation as a group Lasso problem}

It is well-known that the 1-dimensional TV denoising problem (\ref{eq:1Ddenoising}) can be reformulated as a Lasso regression problem by an appropriate change of variable \cite{Harchaoui2008Catching}. We now show that our generalization (\ref{eq:weightedFGL}) can be reformulated as a group Lasso regression problem, which will be convenient for theoretical analysis and implementation \cite{Yuan2006Model}. To this end, we make the change of variables $(\beta,\gamma)\in\RR^{(n-1)\times p} \times \RR^{1\times p}$ given by:
\begin{equation*}
\begin{split}
\gamma &= U_{1,\bullet}\,,\\
\beta_{i,\bullet} &= \frac{U_{i+1,\bullet} - U_{i,\bullet}}{d_i} \quad\text{for }i=1,\ldots,n-1\,.
\end{split}
\end{equation*}
In other words $d_i \beta_{i,j}$ is the jump between the $i$-th and the $(i+1)$-th positions of the $j$-th profile. We immediately get an expression for $U$ as a function of $\beta$ and $\gamma$:
\begin{equation*}
\begin{split}
 U_{1,\bullet} &= \gamma\,,\\
 U_{i,\bullet} &= \gamma + \sum_{j=1}^{i-1} d_j \beta_{j,\bullet} \quad\text{for }i=2,\ldots,n\,.
\end{split}
\end{equation*}
This can be rewritten in matrix form as 
\begin{equation}\label{eq:UfromBeta}
U = \II_{n,1} \gamma + X \beta\,,
\end{equation}
where $X$ is the $n\times (n-1)$ matrix with entries $X_{i,j} = d_j$ for $i>j$, and $0$ otherwise. Making this change of variable, we can re-express \eq{eq:weightedFGL} as follows:
\begin{equation}
\label{eq:reexpress}
\min_{\beta\in\RR^{(n-1)\times p}\,,\gamma\in\RR^{1\times p}} \frac{1}{2} \nm{Y - X\beta -  \II_{n,1} \gamma}^2 + \lambda\sum_{i=1}^{n-1} \nm{\beta_{i,\bullet}}\,.
\end{equation}
For any $\beta\in\RR^{(n-1)\times p}$, the minimum in $\gamma$ is attained with $\gamma = \II_{1,n}(Y-X\beta)/n$. Plugging this into \eq{eq:reexpress}, we get that the matrix of jumps $\beta$ is solution of
\begin{equation}
\label{eq:reexpress2}
\min_{\beta\in\RR^{(n-1)\times p}} \frac{1}{2} \nm{\bar{Y} - \bar{X}\beta}^2 + \lambda\sum_{i=1}^{n-1} \nm{\beta_{i,\bullet}}\,,
\end{equation}
where $\bar{Y}$ and $\bar{X}$ are obtained from $Y$ and $X$ by centering each column.

Equation (\ref{eq:reexpress2}) is now a classical group Lasso regression problem \cite{Yuan2006Model}, with a specific design matrix $\bar{X}$ and groups of features corresponding to the rows of the  matrix $\beta$. The solution $\beta$ of (\ref{eq:reexpress2}) is related to the solution $U$ of our initial problem (\ref{eq:weightedFGL}) by equation (\ref{eq:UfromBeta}).

\section{Implementation}\label{sec:implementation}

Although (\ref{eq:weightedFGL}) and (\ref{eq:reexpress2}) are convex optimization problems that can in principle be solved by general-purpose solvers \cite{Boyd2004Convex}, we want to be able to work in dimensions that reach millions or more, making this computationally difficult. In particular, the design matrix $\bar{X}$ in (\ref{eq:reexpress2}) is a non-sparse matrix of size $n\times (n-1)$, and cannot even fit in a computer's memory when $n$ is large. Moreover, we would ideally like to obtain solutions for various values of $\lambda$, corresponding to various numbers of change-points, in order to be able to select the optimal number of change-points using statistical criteria.  In the single profile case ($p=1$), fast implementations in $O(nk)$ or $O(n\ln n)$ have been proposed  \cite{Friedman2007Pathwise,Harchaoui2008Catching,Hoefling2009path}. However, none of these methods is applicable directly to the $p>1$ setting since they all rely on specific properties of the $p=1$ case, such as the fact that the solution is piecewise-affine in $\lambda$ and that the set of change-points is monotically decreasing with $\lambda$.

In this section we propose two algorithms to respectively \emph{exactly} or \emph{approximately} solve (\ref{eq:weightedFGL}) efficiently. We adopt the algorithms suggested by \cite{Yuan2006Model} to solve the group Lasso problem (\ref{eq:reexpress2}) and show how they can be implemented very efficiently in our case due to the particular structure of the regression problem. We have placed in Annex A several technical lemmas which show how to efficiently perform several operations with the given design matrix $\bar{X}$  that will be used repeatedly in the implementations proposed below.

\subsection{Exact solution by block coordinate descent \label{sec:gLassoimplementation}}
A first possibility to solve the group Lasso problem (\ref{eq:reexpress2}) is to follow a block coordinate descent approach, where each group is optimized in turn with all other groups fixed. It can be shown that this strategy converges to the global optimum, and  is reported to be stable and efficient \cite{Yuan2006Model,Fu1998Penalized}. As shown by \cite{Yuan2006Model}, it amounts to iteratively applying the following equation to each block $i=1,\ldots,n-1$ in turn, until convergence:
\begin{equation}\label{eq:softthreshold}
\beta_{i,\bullet} \leftarrow \frac{1}{\gamma_i}\br{1 - \frac{\lambda}{\nm{S_i}}}_+ S_i\,,
\end{equation}
where $\gamma_i = \nm{ \bar{X}_{\bullet,i}}^2 = i(n-i) d_i^2/n$ and $S_i = \bar{X}_{\bullet,i}^\top \br{\bar{Y} - \bar{X} \beta^{-i}}$, and where $\beta^{-i}$ denotes the $(n-1)\times p$ matrix equal to $\beta$ except for the $i$-th row $\beta^{-i}_{i,\bullet} = 0$. The convergence of the procedure can be monitored by the Karush-Kuhn-Tucker (KKT) conditions:
\begin{equation}\label{eq:kkt}
\begin{split}
-\bar{X}_{\bullet,i}^\top \br{\bar{Y} - \bar{X}\beta} + \frac{\lambda \beta_{i,\bullet}}{\nm{\beta_{i,\bullet}}} = 0
\quad& \forall \beta_{i,\bullet} \neq 0\,,\\
\nm{-\bar{X}_{\bullet,i}^\top \br{\bar{Y} - \bar{X}\beta}} \leq \lambda \quad & \forall \beta_{i,\bullet} = 0\,.
\end{split}
\end{equation}
Since the number of blocks  $n$ can be very large and we expect only a fraction of non-zero blocks at the optimum (corresponding to the change-points), we implemented this block coordinate descent with an active set strategy. In brief, a set of active groups $\Acal$ corresponding to non-zero groups is maintained, and the algorithm alternates between optimizing $\beta$ over the active groups in $\Acal$ and updating $\Acal$ by adding or removing groups based on violation of the KKT conditions. The resulting pseudo-code is shown in Algorithm \ref{algo:grouplasso}. The inner loop (lines 3-7) corresponds to the optimization of $\beta$ on the current active groups, using iteratively block coordinate descent (\ref{eq:softthreshold}). After convergence, groups that have been shrunk to $0$ are removed from the active set (line 8), and the KKT conditions are checked outside of the active set (lines 9-10). If they are not fulfilled, the group that most violates the conditions is added to the active set (line 11), otherwise the current solution satisfies all KKT conditions and is therefore the global optimum (line 13). 

Although it is difficult to estimate the number of iterations needed to reach convergence for a certain level of precision, we note that by Lemma \ref{lem:fastcorrelationwithdesign} (Annex A), computation of $\bar{X}^\top \bar{Y}$ in line 1 can be done in $O(np)$, and each  group optimization iteration (lines 3-7) requires computing $\bar{X}_{\bullet,i}^\top X_{\bullet,\Acal}$ (line 5), done in $O(|\Acal|)$  (see Lemma \ref{lem:XtX} in Annex A), then computing $S_i$ (line 5) in $O(|\Acal| p )$ and soft-thresholding (line 6) in $O(p)$. The overall complexity of each group optimization iteration is therefore $O(|\Acal|p)$. Since each group in $\Acal$ must typically be optimized several times, we expect  complexity that is at least quadratic in $|\Acal|$ and linear in $p$ for each optimization over an active set $\Acal$ (lines 3-7). To check optimality of a solution after optimization over an active set $\Acal$, we need to compute $\bar{X}^\top \bar{X}\beta$ (line 9) which takes $O(np)$ (see Lemma \ref{lem:technical3}, Annex A). Although it is difficult to upper bound the number of iterations needed to optimize over $\Acal$, this shows that a best-case complexity to find $k$ change-points, if we correctly add groups one by one to the active set, would be $O(npk)$ to check $k$ times the KKT conditions and find the next group to add, and $O(p k^3)$ in total if each optimization over an active set $\Acal$ is in $O(p|\Acal|^2)$. In Section \ref{sec:experiments}, we provide some empirical results on the behavior of this block coordinate descent strategy.

\begin{algorithm}
\caption{Block coordinate descent algorithm}
\label{algo:grouplasso}
\begin{algorithmic}[1]
\REQUIRE centered data $\bar{Y}$, regularization parameter $\lambda$.
\STATE Initialize $\Acal \leftarrow \emptyset$, $\beta=0$, $C \leftarrow \bar{X}^\top\bar{Y}$.
\LOOP
\REPEAT
\STATE Pick $i\in\Acal$.
\STATE Compute $S_i \leftarrow C_{i,\bullet} - \bar{X}_{\bullet,i}^\top \bar{X} \beta^{-i}$.
\STATE Update $\beta_{i,\bullet}$ according to (\ref{eq:softthreshold}).
\UNTIL convergence
\STATE Remove inactive groups: $\Acal \leftarrow \Acal \backslash \cbr{i\in\Acal\,:\,\beta_{i,\bullet} = 0}$.
\STATE Check KKT: $S \leftarrow C - \bar{X}^\top \bar{X}\beta$.
\STATE $\hat{u} \leftarrow \argmax_{i\notin A} \nm{S_{i,\bullet}}^2$ , $M=\nm{S_{\hat{u},\bullet}}^2$.
\IF{$M>\lambda^2$}
\STATE Add a new group: $\Acal \leftarrow \Acal\cup\cbr{\hat{u}}$.
\ELSE
\RETURN $\beta$.
\ENDIF
\ENDLOOP
\end{algorithmic}
\end{algorithm}

\subsection{Group fused LARS implementation\label{sec:glarsimplementation}}
Since exactly solving the group Lasso with the method described in Section \ref{sec:gLassoimplementation} can be computationally intensive, it may be of interest to find fast, approximate solutions to (\ref{eq:reexpress2}).
We propose to implement a strategy based on the group LARS, proposed in \cite{Yuan2006Model} as a good way to approximately find the regularization path of the group Lasso. More precisely, the group LARS approximates the solution path of (\ref{eq:reexpress2}) with a piecewise-affine set of solutions and iteratively finds change-points. The resulting algorithm is presented here as Algorithm  \ref{algo:grouplars}, and is intended to approximately solve (\ref{eq:reexpress2}). Change-points are added one by one (lines 4 and 8), and for a given set of change-points the solution moves straight along a descent direction (line 6) with a given step (line 7) until a new change-point is added (line 8). We refer to \cite{Yuan2006Model} for more details and justification for this algorithm.

While the original group LARS method requires storage and manipulation of the design matrix \cite{Yuan2006Model}, implausible for large $n$ here, we can again benefit from the computational tricks provided in Annex A to efficiently run the fast group LARS method. Computing $\bar{X}^\top \bar{Y}$ in line 1 can be done in $O(np)$ using Lemma \ref{lem:fastcorrelationwithdesign}. To compute the descent direction (line 6), we first compute $w$ in $O(|\Acal |p)$ using Lemma \ref{lem:leftmultiplybyinvXAtXA}, then $a$ in $O(np)$ using Lemma \ref{lem:technical3}. To find the descent step (line 7), we need to solve $n$ polynomial equations of degree 2, the coefficients of which are computed in $O(p)$, resulting in a $O(np)$ complexity. Overall the main loop for each new change-point (lines 2--10) takes $O(np)$ in computation and memory, resulting in $O(npk)$ complexity in time and $O(np)$ in memory to find the first $k$ change-points. We provide in Section \ref{sec:experiments} empirical results that confirm this theoretical complexity.

\begin{algorithm}
\caption{Group fused LARS algorithm}
\label{algo:grouplars}
\begin{algorithmic}[1]
\REQUIRE centered data $\bar{Y}$, number of breakpoints $k$.
\STATE Initialize $\Acal\leftarrow\emptyset$, $\hat{c} \leftarrow \bar{X}^\top\bar{Y}$.
\FOR{$i=1$ to $k$}
\IF{i=1}
	\STATE First change-point : $\hat{u} \leftarrow \argmin_{j\in[1,n-1]} \nm{\hat{c}_{j,\bullet}}$, $\Acal \leftarrow \cbr{\hat{u}}$.
	\ENDIF
	\STATE Descent direction: compute $w\leftarrow \br{\bar{X}_{\bullet,\Acal}^\top \bar{X}_{\bullet,\Acal}}^{-1} \hat{c}_{\Acal,\bullet}$ , then $a=\bar{X}^\top  \bar{X}_\Acal w$.
	\STATE Descent step: for each $u\in\sqb{1,n-1}\backslash\Acal$, find if it exists the smallest positive solution $\alpha_u$ of the second-order polynomial in $\alpha$:
$$
\nm{\hat{c}_{u,\bullet} - \alpha a_{u,\bullet}}^2 = \nm{\hat{c}_{v,\bullet} - \alpha a_{v,\bullet}}^2\,,
$$
where $v$ is any element of $\Acal$. 
\STATE Next change-point: $\hat{u} \leftarrow \argmin_{j\in[1,n-1]} \nm{\hat{c}_{j,\bullet}}$, $\Acal \leftarrow \Acal\cup\cbr{\hat{u}}$.
\STATE Update $\hat{c} \leftarrow \hat{c} -\alpha_{\hat{u}} a$.

\ENDFOR

\end{algorithmic}
\end{algorithm}

\section{Theoretical analysis}\label{sec:theory}

In this section, we study theoretically to what extent the estimator (\ref{eq:weightedFGL})  recovers correct change-points. The vast majority of existing theoretical results for offline segmentation and change-point detection consider the setting where $p$ is fixed (usually $p=1$), and $n$ increases (e.g., \cite{Harchaoui2009regularized}). This typically corresponds to cases where we can sample a continuous signal with increasing density, and wish to locate more precisely the underlying change-points as the density increases.

We propose a radically different analysis, motivated notably by applications in genomics. Here, the length of profiles $n$ is fixed for a given technology, but the number of profiles $p$ can increase when more samples or patients are collected. The property we would like to study is then, for a given change-point detection method, to what extent increasing $p$ for fixed $n$ allows us to locate more precisely the change-points. While this simply translates our intuition that increasing the number of profiles should increase the statistical power of change-point detection, and while this property was empirically observed in \cite{Zhang2010Detecting}, we are not aware of previous theoretical results in this setting. In particular we are interested in the consistency of our method, in the sense that it should correctly detect the true change-points if enough samples are available.

\subsection{Consistent estimation of a single change-point}
As a first step towards the analysis of this ``fixed $n$ increasing $p$'' setting, let us assume that the observed centered profiles $\bar{Y}$ are obtained by adding noise to a set of profiles with a \emph{single} shared change-point between positions $u$ and 
$u+1$, for some $u\in[1,n-1]$. In other words, we assume that
$$
\bar{Y} = \bar{X}\beta^* + W\,,
$$
where $\beta^*$ is an $(n-1)\times p$ matrix of zeros except for the $u$-th row $\beta_{u,\bullet}^*$, and $W$ is a noise matrix whose entries are assumed to be independent and identically distributed with respect to a centered Gaussian distribution with variance $\sigma^2$. In this section we study the probability that the first change-point found by our procedure is the correct one, when $p$ increases. We therefore consider  an infinite sequence of jumps $\br{\beta_{u,i}^*}_{i\geq 1}$, and letting $\bar{\beta}^2_p = \frac{1}{p} \sum_{i=1}^p (\beta_{u,i}^*)^2$, we assume that $\bar{\beta}^2 = \lim_{p\rightarrow \infty} \bar{\beta}^2_p $ exists and is finite. We first characterize the first selected change-point as $p$ increases.
\begin{lemma}\label{lem:bp}
Assume, without loss of generality, that $u\geq n/2$, and let, for $i \in \sqb{1,n-1}$,
\begin{equation}\label{eq:gi}
G_i =  d_i^2 \frac{i(n-i)}{n} \sigma^2  +  \frac{\bar{\beta}^2 d_i^2 d_u^2 }{n^2} \times
\begin{cases}
i^2\br{n-u}^2 &\text{ if } i\leq u\,,\\
u^2 \br{n-i}^2 &\text{ otherwise.}
\end{cases}
\end{equation}
When $p\rightarrow +\infty$, the first change-point selected by the group fused Lasso (\ref{eq:weightedFGL}) is in $\argmax_{i\in[1,n-1]} G_i$ with probability tending to $1$.
\end{lemma}

Proof of this result is given in Annex B. From it we easily deduce conditions under which the first change-point is correctly found with increasing probability as $p$ increases. Let us first focus on the unweighted group fused Lasso (\ref{eq:firstproblem}), corresponding to the setting $d_i=1$ for $i=1,\ldots,n-1$.
\begin{theorem}\label{bigtheorem}
Let $\alpha=u/n$ be the position of the change-point scaled in the interval $[0,1]$, and 
\begin{equation}\label{eq:sigmaalpha}
\tilde{\sigma}^2_\alpha = n \bar{\beta}^2 \frac{(1-\alpha)^2(\alpha - \frac{1}{2n})}{\alpha - \frac{1}{2} - \frac{1}{2n}}\,.
\end{equation}
If $\sigma^2 < \tilde{\sigma}^2_\alpha$, the probability that the first change-point selected by the unweighted group fused Lasso (\ref{eq:firstproblem}) is the correct one tends to $1$ as $p\rightarrow +\infty$. When $\sigma^2 > \tilde{\sigma}^2_\alpha$, it is not the correct one with probability tending to $1$.
\end{theorem}

This theorem, the proof of which can be found in Annex C,  deserves several comments.
\begin{itemize}
\item To detect a change-point at position $u=\alpha n$, the noise level $\sigma^2$ must not be larger than the critical value $\tilde{\sigma}^2_\alpha$ given by (\ref{eq:sigmaalpha}), hence the method is not consistent for all positions. $\tilde{\sigma}^2_\alpha$ decreases monotonically from $\alpha=1/2$ to $1$, meaning that change-points near the boundary are more difficult to detect correctly than change-points near the center. The most difficult change-point is the last one ($u=n-1$) which can only be detected consistently if $\sigma^2$ is smaller than
$$
\bar{\sigma}_{1-1/n }^2 = \frac{2 \bar{\beta}^2}{n} + o(n^{-1}) .
$$
\item For a given level of noise $\sigma^2$, change-point detection is asymptotically correct for any $\alpha \in \sqb{\epsilon,1-\epsilon}$, where $\epsilon$ satisfies $\sigma^2 = \tilde{\sigma}^2_{1-\epsilon}$, i.e.,
$$
\epsilon =\sqrt{\frac{\sigma^2}{2 n \bar{\beta}^2}} + o(n^{-1/2})\,.
$$
This shows in particular that increasing the profile length $n$ increases the relative interval (as a fraction of $n$) where change-points are correctly identified, and that we can get as close as we want to the boundary for $n$ large enough.
\item When $\sigma^2 < \tilde{\sigma}^2_\alpha$, the correct change-point is found consistently when $p$ increases, showing the benefit of the accumulation of many profiles.
\end{itemize}

Theorem \ref{bigtheorem} shows that the unweighted group fused Lasso (\ref{eq:firstproblem}) suffers from boundary effects, since it may not correctly identify a single change-points near the boundary is the noise is too large. In fact, Lemma \ref{lem:bp} tells us that if we miss the correct change-point position, it is because we estimate it more towards the middle of the interval (see proof of Theorem \ref{bigtheorem} for details). The larger the noise, the more biased the procedure is. We now show that this issue can be fixed when we consider the weighted group fused Lasso (\ref{eq:weightedFGL}) with well-chosen weights.
\begin{theorem}\label{thm:mainweighted}
The weighted group fused Lasso (\ref{eq:weightedFGL}) with weights given by (\ref{eq:weights}) 
correctly finds the first change-point at any position with probability tending to $1$ as $p\rightarrow +\infty$.
\end{theorem}
The proof of Theorem \ref{thm:mainweighted} is postponed to Annex D. It shows that the weighting scheme (\ref{eq:weights}) cancels the effect of the noise and allows us to consistently estimate any change-point, independently of its position in the signal, as the number of signals increases.

\subsection{Consistent estimation of a single change-point with fluctuating position}
An interesting variant of the problem of detecting a change-point common to many profiles is that of detecting a change-point with similar location in many profiles, allowing fluctuations in the precise location of the change-point. This can be modeled by assuming that the profiles are random, and that the $i$-th profile has a single change-point of value $\beta_i$ at position $U_i$, where $\br{\beta_i,U_i}_{i=1,\ldots,p}$ are independent and identically distributed according to a distribution $P = P_\beta \otimes P_U$ (i.e., we assume $\beta_i$ independent from $U_i$). We denote $\bar{\beta}^2 = E_{P_\beta}\beta^2$ and $p_i = P_U(U=i)$ for $i\in[1,n-1]$. Assuming that the support of $P_U$ is $[a,b]$ with $1\leq a \leq b\leq n-1$, the following result extends Theorem \ref{bigtheorem} by showing that the first change-point discovered by the unweighted group fused Lasso is in the support of $P_U$ under some condition on the noise level, while the weighted group fused Lasso correctly identifies a change-point in the support of $P_U$ asymptotically without conditions on the noise.
\begin{theorem}\label{thm:fluctuate}
\begin{enumerate}
\item 
Let $\alpha=U/n$ be the random position of the change-point on $[0,1]$ and $\alpha_m = a/n$ and $\alpha_M = b/n$ the position of the left and right boundaries of the support of $P_U$ scaled to $[0,1]$. If $1/2\in(\alpha_m,\alpha_M)$, then for any noise level $\sigma^2$, the probability that the first change-point selected by the unweighted group fused Lasso (\ref{eq:firstproblem}) is in the support of $P_U$ tends to $1$ as $p\rightarrow +\infty$. If $1/2 < \alpha_m$ or $\alpha_M < 1/2$, let
\begin{equation}\label{eq:sigmaalpha2}
\tilde{\sigma}^2_{P_U} = n \bar{\beta}^2 \sqb{(1-E\alpha)^2 + \text{var}(\alpha)^2}\times
\begin{cases}
\frac{\alpha_m - \frac{1}{2n}}{\alpha_m - \frac{1}{2} - \frac{1}{2n}} & \text{ if }\alpha_m > \frac{1}{2}\,, \\
\frac{1 - \frac{1}{2n} - \alpha_M }{ \frac{1}{2} - \alpha_M - \frac{1}{2n}} &  \text{ if }\alpha_M < \frac{1}{2}\,. 
\end{cases}
\end{equation}
The probability that the first selected change-point is in the support of $P_U$ tends to $1$ when $\sigma^2 < \tilde{\sigma}^2_{P_U}$. When $\sigma^2 > \tilde{\sigma}^2_{P_U}$, it is outside of the support of $P_U$ with probability tending to $1$. 
\item The weighted group fused Lasso (\ref{eq:weightedFGL}) with weights given by (\ref{eq:weights}) finds the first change-point in the support of $P_U$ with probability tending to $1$ as $p\rightarrow +\infty$, independently of $\sigma^2$ and of the support of $P_U$.
\end{enumerate}
\end{theorem}
This theorem, the proof of which is postponed to Annex E, illustrates the robustness of the method to fluctuations in the precise position of the change-point shared between profiles. Although this situation rarely occurs when we are considering classical multidimensional signals such as financial time series or video signals, it is likely to be the rule when we consider profiles coming from different biological samples, where for example we can expect frequent genomic alterations at the vicinity of  important oncogenes or tumor suppressor genes. Although the theorem only gives a condition on the noise level to ensure that the selected change-point lies in the support of the distribution of change-point locations, a precise estimate of the location of the selected change-point as a function of $P_U$, which generalizes Lemma \ref{lem:bp}, is given in the proof.

\subsection{The case of multiple change-points}\label{sec:conjecture}
While the theoretical results presented above focus on the detection of a single change-point, the real interest of the method is to estimate multiple change-points. The extension of Theorem \ref{bigtheorem} to this setting is, however, not straightforward and we postpone it for future efforts. We conjecture that the group fused Lasso estimator can, under certain conditions, consistently estimate multiple change-points. More precisely, in order to generalize the proof of Theorem \ref{bigtheorem}, we must analyze the path of the vectors $(\hat{c}_{i,\bullet})$, and check that, for some $\lambda$ in (\ref{eq:firstproblem}) or (\ref{eq:weightedFGL}), they reach their maximum norm precisely at the true change-points. The situation is more complicated than in the single change-point case since, in order to fulfill the KKT optimality conditions, the vectors $(\hat{c}_{i,\bullet})$ must hit a hypersphere at each correct change-point, and must remain strictly within the hypersphere between consecutive change-points. This can probably be ensured if the noise level is not too high (like in the single change-point case), and if the positions corresponding to successive change-points on the hypersphere are far enough from each other, which could be ensured if two successive change-points are not too close to each other, and are in sufficiently different directions. Although the weighting scheme (\ref{eq:weights}) ensures consistent estimation of the first change-point independently of the noise level, it may however not be sufficient to ensure consistent estimation of subsequent change-points.

Although we propose no theoretical results besides these conjectures for the case of multiple change-points, we provide experimental results below that confirm that, when the noise is not too large, we can indeed correctly identify several change-points, with probability of success increasing to $1$ as $p$ increases.

\subsection{Estimating the number of change-points\label{dp}}
The number of change-points detected by the group fused Lasso  in the multidimensional signal depends on
the choice of  $\lambda$ in (\ref{eq:firstproblem}) and (\ref{eq:weightedFGL}). In practice, we propose the following scheme in order to estimate a segmentation and the number of change-points. We try to select a $\lambda$ that over-segments the multidimensional
signal, that is, finds more change-points that we would normally expect for the given type of signal or application.
Then, on the set of $k$ change-points found, we perform post-processing using a simple least-squares criteria. 
Briefly,
for each given subset of $k' \leq k$ change-points, we approximate each signal between successive change-points with the
mean value of the points in that interval; then, we calculate the total sum of squared errors (SSE) between the
set of real signals and these piecewise-constant approximations to them. Though it may appear computationally
intensive or even impossible to do this for all subsets of $k' \leq k$ change-points, a dynamic programming
strategy (e.g., \cite{Picard2005statistical}) means that the
best subset of $k' \leq k$ change-points can be calculated for all $k' \in \{1,\ldots,k\}$  in $O(k^3)$.

It then remains to choose the ``best'' $k' \in \{1,\ldots,k\}$ using, for example, a model-selection strategy. The optimal
SSE for $k' + 1$ (which we may call $SSE(k'+1)$ to ease notation), will be smaller than $SSE(k')$ but at a certain point, adding a further change-point will
have no physical reality, it only improves the SSE due to random noise. Here, we implemented a method proposed
in \cite{Lavielle2005Using,Picard2005statistical} where we first normalize the SSE for $k'=1,\ldots,k$ into a score $J(k')$ such that $J(1)=k$ and $J(k)=1$, in such a way the $J(k')$ has an average slope of $-1$ between $1$ and $k$; we then try to detect a kink in the curve by calculating the discrete second derivative of $J(k')$, and selecting the $k'$ after which  this
second derivative no longer rises above a fixed threshold (typically $0.5$).

\section{Experiments}\label{sec:experiments}
In this section we test the group fused Lasso on several simulated and real data sets. All experiments were run under Linux on a
machine
with two 4-core Intel Xeon 3.16GHz  processors and a total of 16Gb of RAM. 
We have implemented the group fused Lasso in MATLAB;
the package \texttt{GFLseg} is available for download\footnote{Available at \url{http://cbio.ensmp.fr/GFLseg}}.

\subsection{Speed trials}
\begin{figure}[htfp]
\begin{center}
\includegraphics[width=0.32\textwidth]{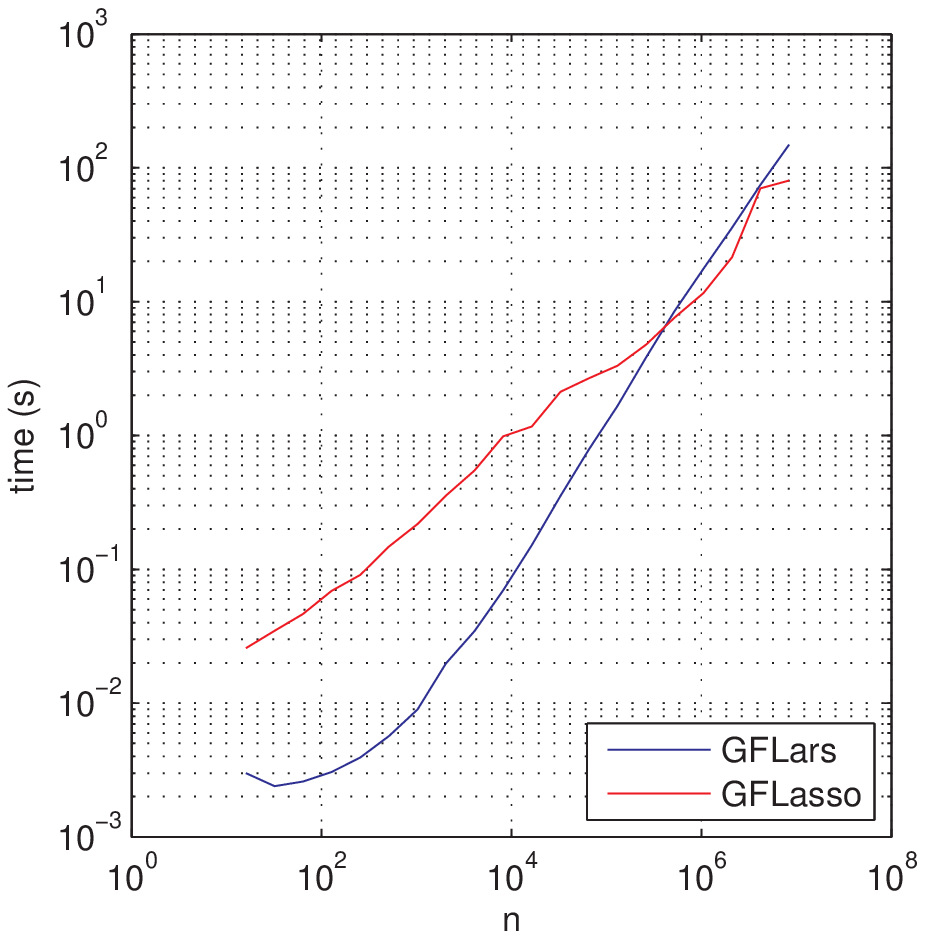}
\includegraphics[width=0.32\textwidth]{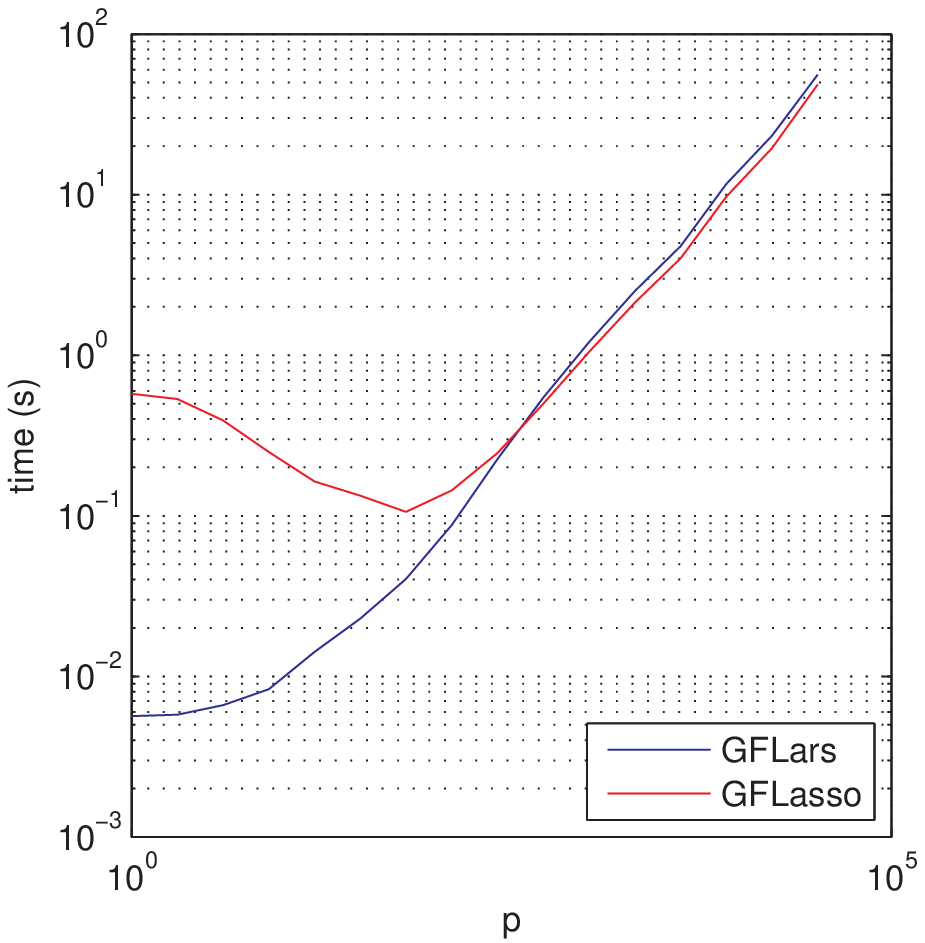}
\includegraphics[width=0.32\textwidth]{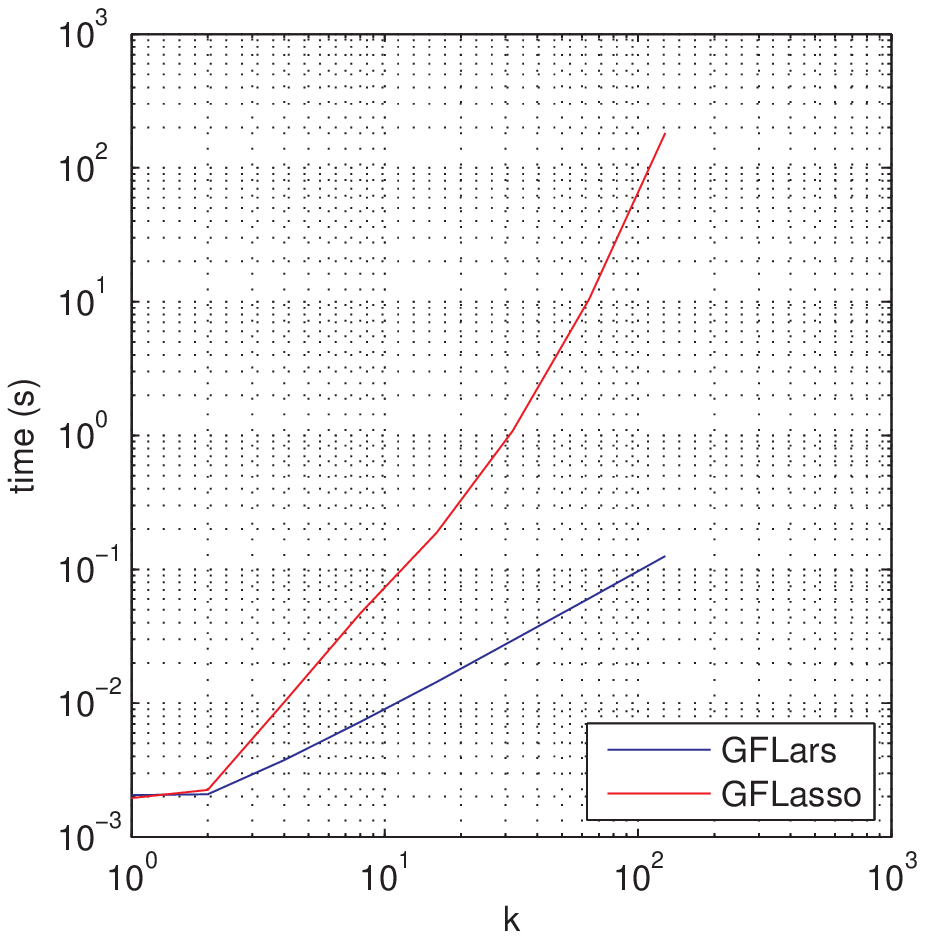}
\end{center}
\caption{\textbf{Speed trials for group fused LARS (top row) and Lasso (bottom row).}  \emph{Left column:} varying $n$, with fixed $p=10$ and $k=10$; \emph{center column:} varying $p$, with fixed $n=1000$ and $k=10$; \emph{right column:} varying $k$, with fixed $n=1000$ and $p=10$. Figure axes are $\log$-$\log$. Results are averaged over 100 trials. 
\label{Fig1}}
\end{figure}

In a first series of experiments, we tested the behavior of the group fused Lasso in terms of computational efficiency. We simulated multidimensional profiles with various lengths $n$ between $2^4$ and $2^{23}$, various dimensions $p$ between $1$ and $2^{15}$, and various number of shared change-points $k$ between $1$ and $2^7$. In each case, we first ran the iterative weighted group fused LARS (Section \ref{sec:glarsimplementation}) to detect successive change-points, and recorded the corresponding $\lambda$ values. We then ran the exact group fused Lasso implementation by block coordinate descent (Section \ref{sec:gLassoimplementation}) on the same $\lambda$ values. Figure \ref{Fig1} shows speed with respect to increasing one of $p$, $n$ and $k$ while keeping the other two variables fixed, for both implementations. The axes are $\log$-$\log$, so the slope gives the exponent of the complexity (resp. $n$, $p$ and $k$). 
For the weighted group fused LARS, linearity is clearest for $k$, whereas for $n$ and $p$, the curves are initially sub-linear, then slightly super-linear for extremely large values of $n$ and $p$. As these time trials reach out to the practical limits of current technology, we see that this is not critical - on average, even the longest trials here took less than 200 seconds. 
The weighted fused group Lasso results are perhaps more interesting, as it is harder to predict in advance the practical time performance of the algorithm. Surprisingly, when increasing $n$ ($p$ and $k$ fixed) or increasing $p$ ($n$ and $k$ fixed), the group fused Lasso eventually becomes as fast the iterative, deterministic group fused LARS. This suggests that at the limits of current technology, if $k$ is small (say, less than 10), the potentially superior performance of the Lasso version (see later) may not even be punished by a slower run-time with respect to the LARS version. We suggest that this may be due to the Lasso optimization problem becoming relatively ``easier'' to solve when $n$ or $p$ increases, as we observed that the Lasso algorithm converged quickly to its final set of change-points.  The main difference between the Lasso and LARS performance appears when the number of change-points increases: with respective empirical complexities cubic and linear in $k$, as predicted by the theoretical analysis, Lasso is already 1,000 times slower than LARS when we seek 100 change-points.

\subsection{Accuracy for detection of a single change-point}
\begin{figure}[htfp]
\begin{center}
\includegraphics[width=14cm]{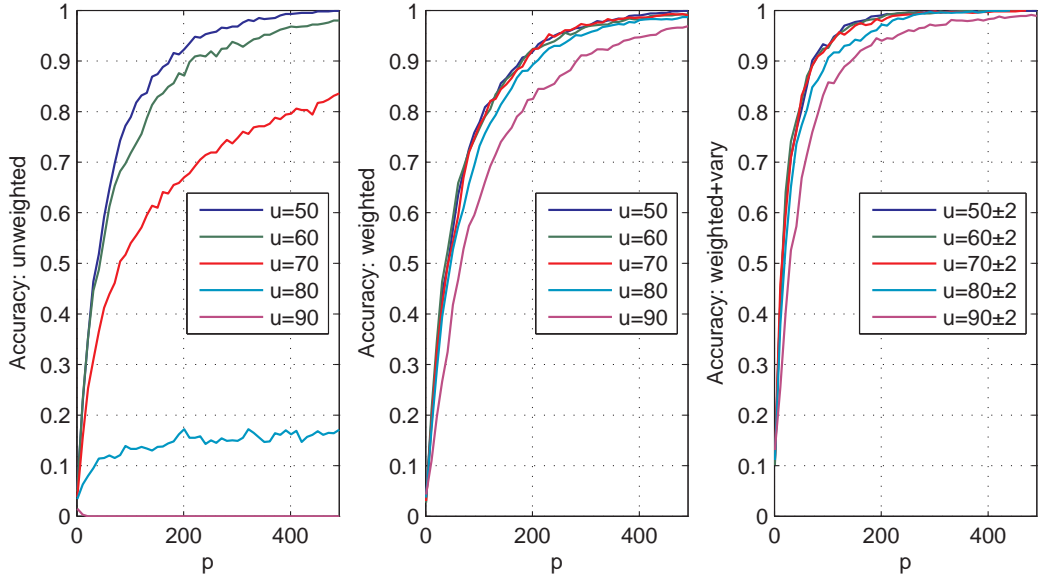}
\end{center}
\caption{\textbf{Single change-point accuracy for the group fused Lasso.}  Accuracy as a function of the number of profiles $p$
when the change-point is placed in a variety of positions  $u =50$ to $u=90$ (left and centre plots, resp. unweighted and
weighted group fused Lasso), or: $u =50 \pm 2$ to $u=90 \pm 2$ (right plot, weighted with varying change-point location), for a signal of length 100.
\label{Fig2}}
\end{figure}

Next, we tested empirically the accuracy the group fused Lasso for detecting a single change-point. We first generated multidimensional profiles of dimension $p$, with a single jump of height $1$ at a position $u$, for different values of $p$ and $u$. We added to the signals an i.i.d. Gaussian noise with variance $\tilde{\sigma}^2_\alpha = 10.78$, the critical value corresponding to $\alpha=0.8$ in Theorem \ref{bigtheorem}. We ran 1000 trials for each value of $u$ and $p$, and recorded how often the group fused Lasso with or without weights correctly identified the change-point. According to Theorem \ref{bigtheorem}, we expect that, for the unweighted group fused Lasso, for $50 \leq u < 80$ there is convergence in accuracy to $1$ when $p$ increases, and for $u > 80$, convergence in accuracy to zero. This is indeed what is seen in Figure \ref{Fig2} (left panel), with $u=80$ the limit case between the two different modes of convergence. The center panel of Figure \ref{Fig2} shows that when the default weights (\ref{eq:weights}) are added,
convergence in accuracy to 1 occurs across all $u$, as predicted by Theorem \ref{thm:mainweighted}.
In addition, the right-hand-side panel of Figure \ref{Fig2} shows results for the same trials except that change-point locations can vary uniformly in the interval $u \pm 2$. We see that, as predicted by Theorem \ref{thm:fluctuate}, the accuracy of the weighted group fused Lasso remains robust against fluctuations in the exact change-point location.

\subsection{Accuracy for detecting multiple change-points}
\begin{figure}[htfp]
\begin{center}
\includegraphics[width=14cm]{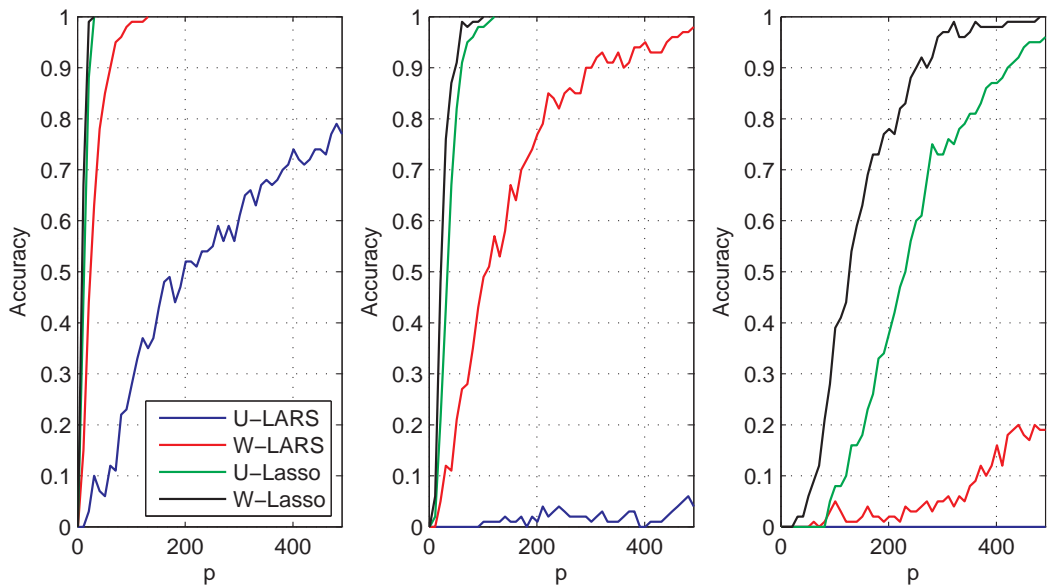}
\end{center}
\caption{\textbf{Multiple change-point accuracy.}  Accuracy as a function of the number of profiles $p$
when change-points are placed at the nine positions $\{10, 20, \ldots, 90\}$ and the variance $\sigma^2$ of the centered Gaussian noise is
either $0.05$ (left), $0.2$ (center) and $1$ (right). The profile length is 100.}
\label{Fig3}
\end{figure}

To investigate the potential for extending the results  to the case of many shared change-points, we further simulated profiles of length $n=100$ with  a change-point at \emph{all} of positions $10, 20, \ldots, 90$. We consider dimensions $p$ between $1$ and $500$. Jumps at
each change-point of each profile were drawn from a Gaussian with mean 0 and variance 1; we then added centered Gaussian noise with $\sigma^2 \in \{0.05,0.2,1\}$ to each position in each profile. For each value of $p$ and $\sigma^2$, we ran one hundred trials of both implementations, with or without weights, and recorded the accuracy of each method, defined as the percentage of trials where the first $9$ change-points detected by the method are exactly the $9$ true change-points. Results are presented in Figure \ref{Fig3} (from left to right, resp. $\sigma^2 = 0.05, 0.2, 1$). Clearly, the group fused Lasso outperforms the group fused LARS, and the weighted version of each algorithm outperforms the unweighted version. Although the group LARS is usually considered a reliable alternative to the exact group Lasso \cite{Yuan2006Model}, this experiment shows that the exact optimization by block coordinate descent may be worth the computational burden if one is interested in accurate group selection. It also demonstrates that, as we conjectured in Section \ref{sec:conjecture}, the group fused Lasso can consistently estimate multiple change-points as the number of profiles increases.

\subsection{Application to gain and loss detection\label{gainloss}}

We now consider a possible application of our method for the detection of regions with frequent gains (positive values) and losses (negative values) among a set of DNA copy number profiles, measured by array comparative genomic hybridization (aCGH) technology \cite{Shah2007Modeling}. We propose a two-step strategy for this purpose: first, find an adequate joint segmentation of the signals; then, check the presence of gain or loss on each interval of the segmentation by summarizing each profile by its average value on the interval. Note that we do not assume that all profiles share exactly the same change-points, but merely see the joint segmentation as an adaptive way to reduce the dimension and remove noise from data.

\begin{figure}[htf]
\begin{center}
\includegraphics[width=10cm]{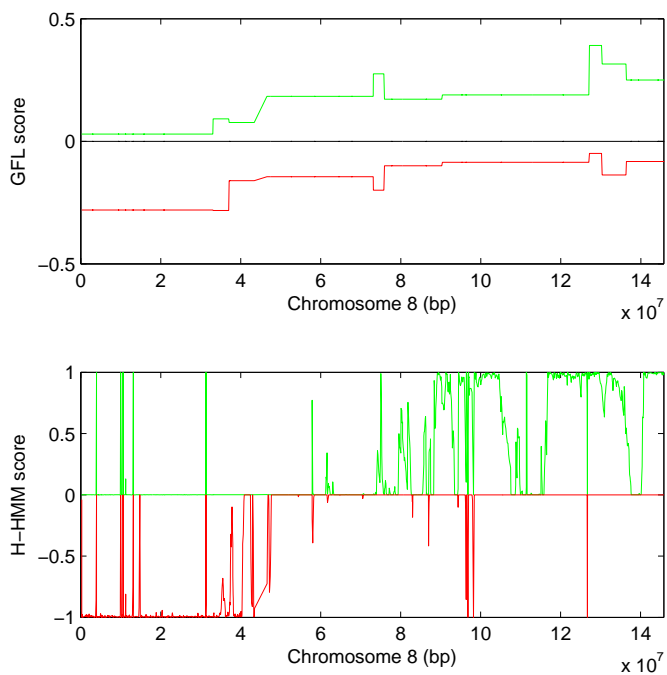}
\end{center}
\caption{\textbf{Joint scores for a set of 18 NSCLC adenocarcinoma cell lines.}  \ref{lung}(a) using weighted group fused LARS; \ref{lung}(b) using H-HMM with the actual code provided by \cite{Shah2007Modeling}.
\label{lung}}
\end{figure}

In practice, we used group fused LARS on each chromosome to identify a set of $100$ candidate change-points, and selected a subset of them by post-processing as described in Section \ref{dp}. Then, in each piecewise-constant interval between successive shared change-points, we calculate the mean of the positive segments (shown in green in Figures \ref{lung}(a) and \ref{piecewise}(c)) and the mean of the negative segments (shown in red). The larger the mean of the positive segments, the more likely we are to believe that a region harbors an important common gain; the reasoning is analogous for important common losses and the mean of the negative segments. Obviously, many other statistical tests could be carried out to detect frequent gains and losses on each segment, once the joint segmentation is performed.

We compare this method for detecting regions of gain and loss with the state-of-the-art H-HMM method \cite{Shah2007Modeling}, which has been shown to outperform several other methods in this setting. As \cite{Shah2007Modeling} have provided their algorithm online with several of their data sets tested in their article, we implemented our method and theirs (H-HMM) on their benchmark data sets. 

\OMIT{For our practical implementation, we used group fused LARS and selected the optimal $k$ using the SSE criteria with dynamic programming
described in Section \ref{dp}. In order to obtain a statistic that summarized well the jointly segmented output of our method and
could help to discover important common gains and losses in aCGH profiles, we proceeded as follows: \emph{i)} run the algorithm, \emph{ii)} in each piecewise-constant interval between successive shared change-points, calculate the mean of the positive segments (shown in green in Figures \ref{lung}(a) and \ref{piecewise}(c)) and the mean of the negative segments (shown in red).
The larger the mean of the positive segments, the more likely we are to believe that a region harbors an important common gain; the reasoning is analagous for important common losses and the mean of the negative segments. }

In the first data set in \cite{Shah2007Modeling}, the goal is to recover two regions -- one amplified, one deleted, that are shared in 8 short profiles, though only 6 of the profiles exhibit each of the amplified or deleted
regions. Performance is measured by area under ROC curve (AUC), following \cite{Shah2007Modeling}. 
Running H-HMM with the default parameters, we obtained
an AUC (averaged over 10 trials) of $0.96 \pm .01$, taking on average 60.20 seconds. The weighted group fused LARS, asked to select 100 breakpoints and followed by dynamic programming, took 0.06 seconds and had
an AUC of $0.97$. Thus, the performance of both methods was similar, though weighted group fused LARS was around 1000 times faster.

The second data set was a cohort of lung cancer cell lines originally published in \cite{Coe2006Differential,Garnis2006High}. As in 
\cite{Shah2007Modeling}, we concentrated on the 18 NSCLC adenocarcinoma (NA) cell lines. Figure \ref{lung} shows
the score statistics obtained on Chromosome 8 when using either weighted group fused LARS or H-HMM. Weighted group fused LARS first selected $100$ candidate change-points per chromosome, then followed optimization of the number of change-points by dynamic programming, took in total 4.7 seconds and finally selected 260 change-points. In contrast,
H-HMM took 38 minutes (100 iterations, as given in the code provided by the authors). The H-HMM scores should look like
those shown in Figure 4 (top panel) of \cite{Shah2007Modeling}; the difference is either due to the stochastic nature of the
algorithm or using a different number of iterations than given in the sample code by the authors. In any case, at the
MYC locus (near $13 \times 10^7$ bp), both methods strongly suggest a common gained region. However, the supposed
advantage of H-HMM to very sparsely predict common gains and losses is not clear here; for example, it gives high common
gain confidence to several fairly large genomic regions between 9 and 14 $\times 10^7$ bp.

\subsection{Application to bladder tumor aCGH profiles}

We further considered a publicly available aCGH data set of 57 bladder 
tumor samples \cite{Stransky2006Regional}. Each aCGH profile gave the 
relative quantity of DNA for 2215 probes. We removed the 
probes corresponding to sex chromosomes because the sex 
mismatch between some patients and the reference made the 
computation of copy number less reliable, giving us a final list of 
2143 probes.

\begin{figure}[htf]
\begin{center}
\includegraphics[width=12cm]{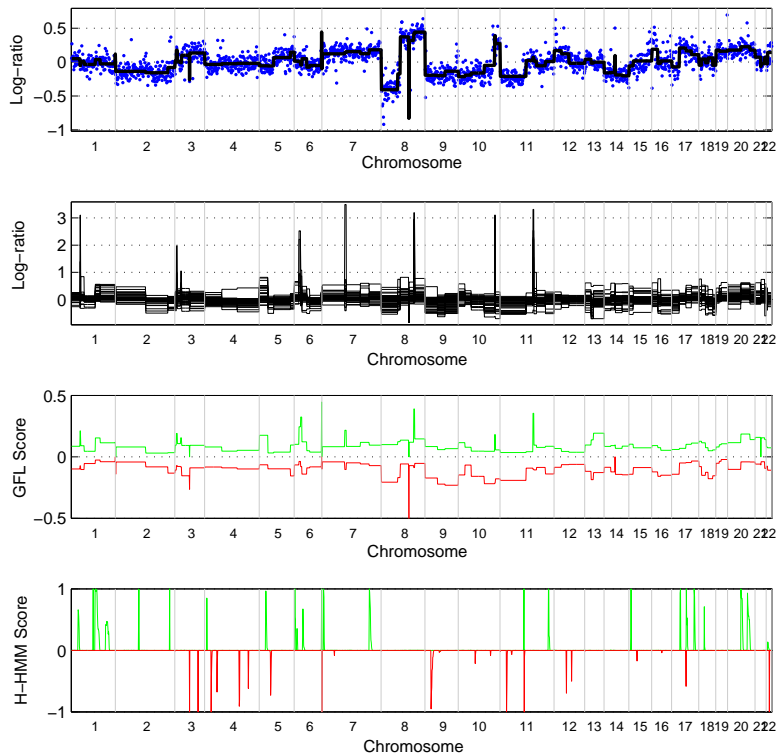}
\end{center}
\caption{\textbf{Bladder cancer profiles.}  \ref{piecewise}(a) shows one of the original 57 profiles and its associated
smoothed version.
\ref{piecewise}(b) shows the result of superimposing the smoothed versions of the 
$57$ bladder tumor aCGH profiles obtained using weighted group fused LARS followed by dimension-selection.
\ref{piecewise}(c) shows the result of transforming the set of smoothed outputs into ``scores'' for amplification/deletion (see Section \ref{gainloss})
and \ref{piecewise}(d) the corresponding output for the H-HMM method \cite{Shah2007Modeling}. Vertical black lines indicate chromosome boundaries.
\label{piecewise}}
\end{figure}

Results are shown in Figure \ref{piecewise}. 97 change-points were selected by the weighted group fused LARS; this  
took 1.1 seconds (Figure \ref{piecewise}(c)). The H-HMM method (Figure \ref{piecewise}(d)) took 13 minutes  for 200 iterations (after
100 iterations convergence had not occured).
We used the comprehensive catalogue of common genomic alterations in bladder cancer provided in Table 2 in \cite{Blaveri2005Bladder} 
to validate the method and compare with H-HMM. Our method (Figure \ref{piecewise}(c))  concurred with the known frequently-amplified chromosome arms 20q, 8q, 19q, 1q, 20p, 17q, 19p, 5p, 2p, 10p, 3q and 7p, and frequently-lost 9p, 9q, 11p, 10q, 13q, 8p, 17p, 18q, 2q, 5q, 18p, 14q and 16q. The only known commonly-lost region which showed unconvincing common loss here was 6q.  As for the H-HMM method (Figure \ref{piecewise}(d)), it selects a small number of very small regions of gain and loss, which are difficult to verify with respect to the well-known frequently amplified arms in 
\cite{Blaveri2005Bladder}. As is suggested, the method may therefore be useful for
selecting the precise location of important genes. However,
as can be seen in Figure \ref{piecewise}(a)-(b), many, but not all, alterations are much larger than those found with H-HMM, and where for
example there are clearly several localized gains and losses in chromosome 8, H-HMM finds nothing at all. Perhaps the complexity of rearrangements in chromosome 8 is not easily taken into account by the H-HMM algorithm. 
Note finally that the weighted group fused LARS
 was more than 700 times faster than H-HMM.

\section{Conclusion}
We have proposed a framework that extends total-variation based approximation to the multidimensional setting, and developed two algorithms to solve the resulting convex optimization problem either exactly or approximately. We have shown theoretically and empirically that the group fused Lasso can consistently estimate a single change-points, and observed experimentally that this property is likely to hold also when several change-points are present. In particular, we observed both theoretically and empirically that increasing the number of profiles is highly beneficial to detect approximatively shared change-points, an encouraging property for biological applications where the accumulation of data measured on cohorts of patients promises to help in the detection of common genomic alterations. 

Although we do not assume that all profiles have the same change-points, we estimate only shared change-points. In other words, we try to estimate the union of the change-points present in the profiles. This can be useful by itself, eg, for dimension reduction. If we wanted to detect change-points of individual profiles, we may either post-process the results of the group fused Lasso, or modify the formulation by, e.g., adding a TV penalty to each profile in addition to the group lasso penalty. Similarly, for some applications, we may want to add a $\ell_1/\ell_2$ norm to the group fused Lasso objective function in order to constrain some or all signals to be frequently null. Finally, from a computational point of view, we have proposed efficient algorithms to solve an optimization problem (\ref{eq:weightedFGL}) which is the proximal operator of more general optimization problems where a smooth convex functional of $U$ is minimized with a constraint on the multidimensional TV penalty; this paves the way to the efficient minimization of such functionals using, e.g., accelerated gradient methods \cite{Beck2009fast}.

\OMIT{
\begin{itemize}
\item A new way to analyze problem ($n$ fixed, $p$ increases), relevant for applications in biology. Some preliminary results. We indeed show theoretically and empirically that increasing $p$ helps a lot.
\item insist on the robustness: we look for commonly (but not always) shared breakpoints with similar (but not same) location
\item conjecture that the theoretical analysis holds for more breakpoints, although more technical. 
\item The effect of increasing $p$ is twofold: concentration of measure kills deviations of the noise, and higher dimensions allows decorrelations of successive change-points which is beneficial (as opposed to staircase effect in 1D)
\item choice of $k$: not the focus here, we suggest to post-process the results like everybody does (add refs).
\item Other implementations to follow exactly the group Lasso solution would be interesting, but it must be fast.
\item Modification of the group Lasso procedure to remove the positional effect would be nice.
\item We have not discussed in detail the problem of choosing the number of change-points, and suggest in practice to use existing criteria for this purpose \cite{Birge2001Gaussian,Lavielle2006Detection}. It should be noted, however, that formulating the change-point detection problem as a group Lasso regression problem may suggest specific strategies..?
\end{itemize}
}


\section*{Annex A: Computational lemmas}
In this Annex we collect a few results useful to carry out the fast implementations claimed in Section \ref{sec:implementation}. Remember that the $n\times (n-1)$ matrix $X$ defined in (\ref{eq:UfromBeta}) is defined by $X_{i,j} = d_j$ for $i>j$, $0$ otherwise. Since the design matrix $\bar{X}$ of the group Lasso problem (\ref{eq:reexpress2}) is obtained by centering each column of $X$ to zero mean, its columns are given by:
\begin{equation}\label{eq:xbar}
\forall i=1,\ldots, n-1, \quad \bar{X}_{\bullet,i} = \br{\underbrace{\br{\frac{i}{n}-1}d_i,\ldots,\br{\frac{i}{n}-1}d_i}_i , \underbrace{\frac{i}{n}d_i,\ldots,\frac{i}{n}d_i}_{n-i}}^\top\,.
\end{equation}

We first show how to compute efficiently $\bar{X}^\top R$ for any matrix $R$:
\begin{lemma}\label{lem:fastcorrelationwithdesign}
For any $R\in\RR^{n\times p}$, we can compute $C = \bar{X}^\top R$ in $O(np)$ operations and memory as follows:
\begin{enumerate}
\item Compute the $n\times p$ matrix $r$ of cumulative sums $r_{i,\bullet} = \sum_{j=1}^i R_{j\bullet}$ by the induction:
\begin{itemize}
\item $r_{1,\bullet} = R_{1,\bullet}$ .
\item For $i=2,\ldots,n$, $r_{i,\bullet} = r_{i-1,\bullet} + R_{i,\bullet}$ .
\end{itemize}
\item For $i=1,\ldots, n-1$, compute $C_{i,\bullet} = d_i \br{i r_{n,\bullet} / n - r_{i,\bullet}}$ .
\end{enumerate}
\end{lemma}
\begin{proof}
Using (\ref{eq:xbar}) we obtain the $i$-th row of $C=\bar{X}^\top R$, for $i=1,\ldots,n-1$, as follows:
\begin{equation*}
\begin{split}
C_{i,\bullet} & = \bar{X}_{\bullet,i}^\top R\\
& = \br{\frac{i}{n}-1} d_i \br{\sum_{j=1}^i R_{j,\bullet}} + \frac{i}{n}  d_i \br{\sum_{j=i+1}^n R_{j,\bullet}} \\
&= d_i\br{\frac{i}{n} r_{n,\bullet} - r_{i,\bullet}}\,.
\end{split} 
\end{equation*}
\end{proof}

Next, we show how to compute efficiently submatrices of the $(n-1)\times(n-1)$ matrix $\bar{X}^\top \bar{X}$.
\begin{lemma}\label{lem:XtX}
For any two subsets of indices $A=\br{a_1,\ldots,a_{|A|}}$ and $B=\br{b_1,\ldots,b_{|B|}}$ in $[1,n-1]$, the matrix $\bar{X}^\top_{\bullet,A} \bar{X}_{\bullet,B}$ can be computed in $O\br{|A||B|}$ in time and memory with the formula:
\begin{equation}
\forall (i,j)\in\sqb{1,|A|}\times\sqb{1,|B|}\,,\quad \sqb{\bar{X}^\top_{\bullet,A} \bar{X}_{\bullet,B}}_{i,j} = d_{a_i} d_{b_j} \frac{\min(a_i,b_j) \sqb{n-\max(a_i,b_j)}}{n}\,.
\end{equation}
\end{lemma}
\begin{proof}
Let us denote $V = \bar{X}_{\bullet,A}^\top \bar{X}_{\bullet,B}$. For any $(i,j)\in\sqb{1,|A|}\times\sqb{1,|B|}$, denoting $u=\min(a_i,b_j)$ and $v=\max(a_i,b_j)$, we easily get from \eq{eq:xbar} an explicit formula for $V_{i,j}$, namely,
\begin{equation*}
\begin{split}
V_{i,j} &= \bar{X}_{\bullet,{a_i}}^\top \bar{X}_{\bullet,{b_j}}\\
&= d_u d_v\sqb{u\br{\frac{u}{n}-1}\br{\frac{v}{n}-1} + (v-u)\frac{v}{n}\br{\frac{u}{n}-1} + (n-v)\frac{u}{n}\frac{v}{n}}\\
&= d_ud_v\frac{u(n-v)}{n}\,.
\end{split}
\end{equation*}
\end{proof}

The next lemma provides another useful computational trick to compute efficiently $\bar{X}^\top \bar{X} R$ for any matrix $R$:
\begin{lemma}\label{lem:technical3}
For any $R \in\RR^{(n-1)\times p}$, we can compute $C=\bar{X}^\top \bar{X} R$ in $O(np)$ by
\begin{enumerate}
\item Compute, for $i=1,\ldots,n-1$, $\tilde{R}_{i,\bullet} = d_i R_{i,\bullet}$.
\item Compute the $1\times p$ vector $S=\br{\sum_{i=1}^{n-1}i \tilde{R}_{i,\bullet}}/n$.
\item Compute the $(n-1)\times p$ matrix $T$ defined by $T_{i,\bullet} = \sum_{j=i}^{n-1}\tilde{R}_{j,\bullet}$ by the induction:
\begin{itemize}
\item $T_{n-1,\bullet} = \tilde{R}_{n-1,\bullet}$.
\item for $i=n-2,\ldots, 1$, $T_{i,\bullet} = T_{i+1,\bullet} + \tilde{R}_{i,\bullet}$.
\end{itemize}
\item Compute the $(n-1)\times p$ matrix $U$ defined by $U_{i,\bullet} = \sum_{j=1}^{i}\br{S - T_{j,\bullet}}$ by the induction:
\begin{itemize}
\item $U_{1,\bullet} = S-T_{1,\bullet}$.
\item for $i=2,\ldots, n-1$, $U_{i,\bullet} = U_{i-1,\bullet} + S-T_{i,\bullet}$.
\end{itemize}
\item Compute, for $i=1,\ldots,n-1$, $C_{i,\bullet} = d_i U_{i,\bullet}$
\end{enumerate}
\end{lemma}
Each step in Lemma \ref{lem:technical3} has complexity $O(np)$ in memory and time, leading to an overall complexity in $O(np)$ to compute $\bar{X}^\top \bar{X} R$. We note that if $R$ is row-sparse, i.e., is several rows of $R$ are null, then the first two steps have complexity $O(sp)$, where $s$ is the number of non-zero rows in $R$. Although this does not change the overall complexity to compute $\bar{X}^\top \bar{X} R$, this leads to a significant speed-up in practice when $s \ll n$.
\begin{proof}
Let us denote $D$ the $(n-1)\times (n-1)$ diagonal matrix with entries $D_{i,i}=d_i$. By Lemma \ref{lem:XtX}, we know that $\bar{X}^\top\bar{X} = DVD$, with $V_{i,j} = \min(i,j)\sqb{n-\max(i,j)}/n$, for $1\leq i,j\leq n-1$. Since step 1 computes $\tilde{R}=DR$ and step 5 computes $C = DU$, we just need to show that the $U$ computed in step 4 satisfies $U=V\tilde{R}$ to conclude that $C=DV\tilde{R}=DVDR=\bar{X}^\top\bar{X} R$. By step 4, $U$ is defined by the relation $U_{i,\bullet} - U_{i-1,\bullet} = S - T_{i,\bullet}$ for $i=1,\ldots,n-1$ (with the convention $U_{0,\bullet}=0$), therefore we just need to show that $\br{V_{i,\bullet} - V_{i-1,\bullet}}\tilde{R} = S - T_{i,\bullet}$ for $i=1,\ldots,n-1$ to conclude. For $0\leq j < i \leq n-1$, we note that $V_{i,j} = j(n-i)/n$ (with the convention $V_{0,\bullet}=0$) and $V_{i-1,j} = j(n-i+1)/n$, and therefore $V_{i,j} - V_{i-1,j} = -j/n$. For $1\leq i \leq j\leq n-1$, we have $V_{i,j} = i(n-j)/n$ and $V_{i-1,j} = (i-1)(n-j)/n$ and therefore $V_{i,j} - V_{i-1,j} = 1-j/n$. Combining these expressions we get, for $i=1,\ldots,n-1$:
\begin{equation*}
\br{V_{i,\bullet} - V_{i-1,\bullet}}\tilde{R} = -\sum_{j=1}^{n-1}\frac{j\tilde{R}_{j,\bullet}}{n} + \sum_{j=i}^{n-1}\tilde{R}_{j,\bullet} = S - T_{i,\bullet}\,,
\end{equation*}
where $S$ and $T$ are defined in steps 2 and 3. This concludes the proof that $C=\bar{X}^\top\bar{X} R$.
\end{proof}

Next we show that $\br{\bar{X}^\top\bar{X}}^{-1}$ has a tridiagonal structure, resulting in fast matrix multiplication.
\begin{lemma}\label{lem:leftmultiplybyinvXAtXA}
For any set $A=\br{a_1,\ldots,a_{|A|}}$ of distinct indices with $1\leq a_1 < \ldots < a_{|A|}\leq n-1$, the matrix $\br{\bar{X}_{\bullet,A}^\top \bar{X}_{\bullet,A} }$ is invertible, and for any $|A|\times p$ matrix $R$, the matrix 
$$
C = \br{\bar{X}_{\bullet,A}^\top \bar{X}_{\bullet,A} }^{-1} R
$$
can be computed in $O(|A|p)$ in time and memory by 
\begin{enumerate}
\item For $i=1,\ldots,|A|-1$, compute 
$$
\Delta_i = \frac{d_{a_{i+1}}^{-1}R_{i+1,\bullet} - d_{a_i}^{-1}R_{i,\bullet}}{a_{i+1}-a_i}\,.
$$ 
\item Compute the successive rows of $C$ according to:
\begin{equation}\label{eq:inversemultiply}
\begin{split}
C_{1,\bullet} &= d_{a_1}^{-1} \br{\frac{R_{1,\bullet}}{a_1} - \Delta_1}\,,\\
C_{i,\bullet} &= d_{a_i}^{-1} \br{\Delta_{i-1} - \Delta_i}\quad \text{for }i=2,\ldots,|A|-1\,,\\
C_{|A|,\bullet} &= d_{a_{|A|}}^{-1} \br{\Delta_{|A|-1}+ \frac{R_{|A|,\bullet}}{n-a_{|A|}}}\,.
\end{split}
\end{equation}
\end{enumerate}
\end{lemma}
\begin{proof}
Let us denote $V = \bar{X}_{\bullet,A}^\top \bar{X}_{\bullet,A}$. By Lemma \ref{lem:XtX} we know that, for $1\leq i \leq j\leq |A|$,
\begin{equation*}
V_{i,j} = d_{a_i} d_{a_j} \frac{{a_i}(n-{a_j})}{n}\,.
\end{equation*}
$V$ being symmetric semi-separable, one can easily check that $V$ is invertible and admits as inverse a tridiagonal matrix with the following entries \cite{Baranger1971Matrices}:
\begin{equation}
\begin{split}
V^{-1}_{i,i} &= d_{a_i}^{-2} \br{\ovr{a_i - a_{i-1}} + \ovr{a_{i+1}-a_i} }\quad\text{for }i=1,\ldots,|A|,\\
V^{-1}_{i,i+1} = V^{-1}_{i+1,i} &= - \frac{d_{a_i}^{-1} d_{a_{i+1}}^{-1}}{a_{i+1} - a_i} \quad\text{for }i=1,\ldots,|A|-1, 
\end{split}
\end{equation}
where by convention we define $a_0=0$ and $a_{|A|+1} = n$. This tri-diagonal structure allows successive rows of $C$ to be
expressed as a sum of just a few terms. More precisely, for $1<i<|A|$, we obtain:
\begin{equation*}
\begin{split}
C_{i,\bullet} &= -\frac{d_{a_{i-1}}^{-1} d_{a_i}^{-1} R_{i-1,\bullet}}{a_i - a_{i-1}} + d_{a_i}^{-2}R_{i,\bullet}\br{ \ovr{a_i - a_{i-1}} + \ovr{a_{i+1}-a_i} } - \frac{d_{a_i}^{-1} d_{a_{i+1}}^{-1} R_{i+1,\bullet}}{a_{i+1} - a_i}\\
&= d_{a_i}^{-1} \br{\frac{d_{a_i}^{-1} R_{i,\bullet} - d_{a_{i-1}}^{-1}R_{i-1,\bullet}}{a_i - a_{i-1}} + \frac{d_{a_i}^{-1} R_{i,\bullet} - d_{a_{i+1}}^{-1} R_{i+1,\bullet}}{a_{i+1} - a_i}}\\
&= d_{a_i}^{-1}  \br{\Delta_{i-1} - \Delta_{i}}\,.
\end{split}
\end{equation*}
Similarly, for $i=1$ and $i=|A|$ we easily recover \eq{eq:inversemultiply}.
\end{proof}

\section*{Annex B: Proof of Lemma \ref{lem:bp}}

The solution of (\ref{eq:weightedFGL}) is constant, i.e., corresponds to $\beta=0$ (no change-point), as long as the KKT conditions (\ref{eq:kkt}) are satisfied for $\beta=0$. This translates to $\nm{\bar{X}^\top _{\bullet,i} \bar{Y}} \leq \lambda$ for all $i$. The first change-point occurs when $\lambda = \max_i \nm{\bar{X}^\top _{\bullet,i} \bar{Y}}$, and the change-point is precisely located in the position $i$ that reaches the maximum. Therefore the first change-point is the row with the largest Euclidean norm of the matrix:
$$
\hat{c} = \bar{X}^\top \bar{Y} =  \bar{X}^\top \bar{X}\beta^* +  \bar{X}^\top W\,.
$$
The entries of the matrix $\hat{c}$ are therefore jointly Gaussian. Since only the $u$-th row $\beta_{u,\bullet}$ of $\beta$ is non-zero, we get
$$
E(\hat{c}) = \bar{X}^\top \bar{X}\beta^* = \bar{X}^\top \bar{X}_{\bullet,u} \beta_{u,\bullet}^*\,.
$$
Using Lemma \ref{lem:XtX} we compute:
 \begin{equation}\label{eq:mean}
E(\hat{c}_{i,\bullet}) = \sqb{\bar{X}^\top \bar{X}\beta^*}_{i,\bullet} = 
 \begin{cases}
 d_i d_u \frac{i(n-u)}{n} \beta_{u,\bullet}^* & \text{for }1\leq i\leq u\,,\\
 d_i d_u \frac{u(n-i)}{n}  \beta_{u,\bullet}^* & \text{for }u\leq i\leq n-1\,.
 \end{cases}
 \end{equation}
 On the other hand, by (\ref{eq:xbar}) we have for any $i\in[1,n-1]$,
 $$
\sqb{ \bar{X}^\top W}_{i,\bullet} = d_i \sqb{\sum_{j=1}^i \br{\frac{i}{n} - 1} W_{j,\bullet} + \sum_{j=i+1}^n \frac{i}{n} W_{j,\bullet} }\,.
 $$
 Since
 $$
 E\br{W_{i,\bullet}^\top W_{j,\bullet}} = \delta_{i,j}\sigma^2 \Id_p\,,
 $$
 where $\delta_{i,j}$ is the Dirac function, we have for $1\leq i \leq j \leq n-1$:
 \begin{equation}\label{eq:cov}
 \begin{split}
 E&\br{\sqb{ \bar{X}^\top W}_{i,\bullet}^\top \sqb{ \bar{X}^\top W}_{j,\bullet}  } \\
 &=  d_i d_j \sqb{i\br{\frac{i}{n}-1}\br{\frac{j}{n}-1} + \br{j-i}\frac{i}{n}\br{\frac{j}{n} - 1} + \br{n-j}\frac{i}{n}\frac{j}{n} }  \sigma^2\Id_p\\
 &= d_i d_j \frac{i\br{n-j}}{n} \sigma^2\Id_p\,.
 \end{split}
 \end{equation}
 In summary, we have shown that $\hat{c}$ is jointly Gaussian with $E\br{\hat{c}_{i,\bullet}}$ given by \eq{eq:mean} and covariance between $\hat{c}_{i,\bullet}$ and $\hat{c}_{j,\bullet}$ given by \eq{eq:cov}.
 
 In particular, if we denote
 $
 F_i = \nm{\hat{c}_{i,\bullet}}^2 \,,
 $
 then, for $i\leq u$, $F_i n / \br{d_i^2 i(n-i) \sigma^2}$ follows a non-central $\chi^2$ distribution with $p$ degrees of freedom and non-centrality parameter $p \bar{\beta}_p^2 d_u^2 i (n-u)^2 / \sqb{n(n-i)\sigma^2}$. In particular,
$$
E F_i =  p \bar{\beta_p}^2  d_i^2 d_u^2 \frac{i^2\br{n-u}^2}{n^2}  + p d_i^2 \frac{i(n-i)}{n} \sigma^2\, ,
$$
and since $\lim_{p\rightarrow +\infty} \bar{\beta_p}^2 = \bar{\beta}^2$, we get that $F_i / p$ converges in probability to
\begin{equation}\label{eq:giminus}
G_i = \frac{E F_i}{p} = \bar{\beta}^2 d_i^2 d_u^2 \frac{i^2\br{n-u}^2}{n^2}  + d_i^2 \frac{i(n-i)}{n} \sigma^2 \,.
\end{equation}
A similar computation shows that for $i\geq u$, $F_i / p$ converges in probability to
\begin{equation}\label{eq:giplus}
G_i = \bar{\beta}^2 d_i^2 d_u^2 \frac{u^2\br{n-i}^2}{n^2}  + d_i^2 \frac{i(n-i)}{n} \sigma^2 \,.
\end{equation}
Note that (\ref{eq:giminus}) and (\ref{eq:giplus}) are equivalently defined in (\ref{eq:gi}). Now, let $V = \argmax_{i\in [1,n-1]}G_i$. For any $v\in V$ and $j \notin V$, the probability of the event $F_v>F_j$ tends to $1$, because $G_v > G_j$. By the union bound the probability of the event $\max_{j\notin V} F_i < \max_{v\in V} F_v$ also converges to $1$, showing that the probability to select a change-point in $V$ converges to $1$ as $p\rightarrow +\infty$.\qed

\OMIT{
Let us assume without loss of generality that $u\geq n/2$. Then $G_i$ is decreasing on $[u,n-1]$, as it is a sum of two decreasing functions of $i$ on this interval, and therefore 
$$
\max_{i\in[1,n]} G_i = \max_{i\in[1,u]} G_i \,.
$$
To conclude the proof it suffices to observe that, assuming the uniqueness of
\begin{equation}
\hat{u} = \argmax_{i\in[1,n]} G_i =  \argmax_{i\in[1,u]} \bar{\beta}^2 \frac{i^2 \br{n-u}^2}{n^2} + \sigma^2 \frac{i\br{n-i}}{n}\,,
\end{equation}
the probability of the event $F_{\hat{u}} > F_i$ tends to $1$ for any $i\neq \hat{u}$ as $p\rightarrow +\infty$ because $G_{\hat{u}} > G_i$, and by the union bound the probability of the event $F_{\hat{u}} = \max_{i\in[1,u]} F_i$ also converges to $1$. This shows that the probability to select $\hat{u}$ converges to $1$ as $p\rightarrow +\infty$.\qed
}

\section*{Annex C: Proof of Theorem \ref{bigtheorem}}
By Lemma \ref{lem:bp}, we know that the first change-point selected by (\ref{eq:weightedFGL}) is in  $\argmax_{i\in[1,n]} G_i$ with probability tending to $1$ as $p$ increases, where $G_i$ is defined in (\ref{eq:gi}). We will therefore asymptotically select the correct change-point $u$ if and only if $G_u = \max_{i\in [1,n-1]} G_i$. Remember we assume, without lack of generality, that $u\geq n/2$. For $u\leq i \leq n-1$, we observe that $G_i$ given by (\ref{eq:giplus}) is a decreasing function of $i$ as a sum of two decreasing functions. Therefore, it always holds that $G_u = \max_{i\in [u,n-1]} G_i$, and we just need to check whether or not $G_u = \max_{i\in [1,u]} G_i$ holds.

For $i\in[1,u]$, $G_i$ given by (\ref{eq:giminus}) is a second-order polynomial of $i$, which is equal to $0$ at $i=0$ and strictly positive for $i=u$. Therefore $G_u = \max_{i\in [1,u]} G_i$ if and only if $G_u > G_{u-1}$. Let us therefore compute:
\begin{equation}
\begin{split}
G_u - G_{u-1}
&=\bar{\beta}^2\frac{ (n-u)^2}{n^2}\sqb{u^2 - (u-1)^2}+ \frac{\sigma^2 }{n} \sqb{u(n-u) -(u-1)(n-u+1)}\\
&= \frac{{\bar{\beta}}^2 (2u-1)(n-u)^2}{n^2} + \frac{\sigma^2 (n-2u+1)}{n}\\
&= 2 \sqb{ \bar{\beta}^2 n \br{1 - \alpha}^2 \br{\alpha - \frac{1}{2n}}  + \sigma^2 \br{\frac{1}{2}-\alpha + \frac{1}{2n}}}\\
&= 2 \br{\tilde{\sigma}^2 - \sigma^2 } \br{\alpha - \frac{1}{2} - \frac{1}{2n}}\,,
\end{split}
\end{equation}
where $\alpha=u/n$ and 
$$
\tilde{\sigma}^2 = n \bar{\beta}^2 \frac{(1-\alpha)^2(\alpha - \frac{1}{2n})}{\alpha - \frac{1}{2} - \frac{1}{2n}}\,.
$$
This shows that, when $\alpha > 1/2 + 1/(2n)$, $G_u > G_{u-1}$ if and only if $\sigma < \tilde{\sigma}$. On the other hand, when $\alpha = 1/2$ or $1/2 + 1/(2n)$, we have always that $G_u > G_{u-1}$.\qed

\section*{Annex D: Proof of Theorem \ref{thm:mainweighted}}
As for the proof of Theorem \ref{bigtheorem}, we need to check whether or not $G_u = \max_{i\in[1,n-1]}G_i$, where $G_i$ is defined in (\ref{eq:gi}), to deduce whether the method selects the correct change-point $u$ or a different position with probability tending to $1$ when $p$ increases. Substituting weights $d_i$ defined in (\ref{eq:weights}) into $G_i$, we obtain:
\begin{equation}\label{eq:annexD}
G_i =   \sigma^2  +  \bar{\beta}^2 \times
\begin{cases}
i\br{n-u}/u\br{n-i} &\text{ if } i\leq u\,,\\
u\br{n-i}/i\br{n-u}  &\text{ otherwise.}
\end{cases}
\end{equation}
It is then easy to see that (\ref{eq:annexD}) is increasing on $[1,u]$, and decreasing on $[u,n-1]$, showing that we always have $\argmax_{i\in[1,n-1]} G_i = u$. The result then follows from Lemma \ref{lem:bp}. \qed

\section*{Annex E: Proof of Theorem \ref{thm:fluctuate}}
Following the proof of Lemma \ref{lem:bp}, let us estimate $F_i = \nm{\hat{c}_{i,\bullet}}^2$ for $i\in[1,n-1]$. For any $j\in[1,p]$, we first observe by (\ref{eq:mean}) that
 \begin{equation}\label{eq:mean2}
 \sqb{\bar{X}^\top \bar{X}\beta}_{i,j} = 
 \begin{cases}
d_i d_{U_j} \frac{i(n-U_j)}{n} \beta_{j} & \text{if } i \leq U_j\,,\\
d_i d_{U_j} \frac{U_j(n-i)}{n}  \beta_{j} & \text{otherwise}\,.
 \end{cases}
 \end{equation}
Therefore,
\begin{equation}
\sum_{j=1}^p  \sqb{\bar{X}^\top \bar{X}\beta}_{i,j} ^2 = \frac{d_i^2}{n^2} \sum_{j=1}^p d_{U_j}^2 \beta_j^2 \sqb{i^2 \br{n-U_j}^2 \II(i \leq U_j) + \br{n-i}^2 U_j^2  \II(i > U_j)}\,,
\end{equation}
and by independence of $\beta_i$ and $U_i$:
$$
\frac{1}{p} E\sum_{j=1}^p  \sqb{\bar{X}^\top \bar{X}\beta}_{i,j} ^2  = \bar{\beta}^2 \frac{d_i^2}{n^2} \sqb{\sum_{u=1}^i p_u d_u^2 u^2 (n-i)^2 + \sum_{u=i+1}^{n-1} p_u d_u^2 \br{n-u}^2 i^2}\,.
$$
Since $(\beta_i,U_i)_{i=1,\ldots,p}$ are independent of the noise, we obtain that $F_i/p$ converges in probability to
\begin{equation}\label{eq:Ginoise}
G_i = \bar{\beta}^2 \frac{d_i^2}{n^2} \sqb{\sum_{u=1}^i p_u d_u^2 u^2 (n-i)^2 + \sum_{u=i+1}^{n-1} p_u d_u^2 \br{n-u}^2 i^2} + d_i^2 \frac{i(n-i)}{n}\sigma^2\,.
\end{equation}
As in Lemma \ref{lem:bp} we can conclude that the method will select the position
$$
\hat{u} = \argmax_{u\in[1,n-1]} G_i
$$
with probability tending to 1 as $p$ increases.

Let us now assume that the support of $P_U$ is an interval $[a,b]$ (corresponding to a possible range of fluctuation of a change-point). Then, we observe that for $i\leq a$, $G_i$ in (\ref{eq:Ginoise}) reduces to
\begin{equation}\label{eq:Ginoisefirstterm}
\begin{split}
G_i &= \bar{\beta}^2 \frac{d_i^2}{n^2} \sqb{ 0 + \sum_{u=a}^{b} p_u d_u^2 \br{n-u}^2 i^2} + d_i^2 \frac{i(n-i)}{n}\sigma^2 \\
&= \bar{\beta}^2 \frac{i^2 d_i^2 }{n^2} E\sqb{d_U(n-U)}^2 + d_i^2 \frac{i(n-i)}{n}\sigma^2\,.
\end{split}
\end{equation}

Let us now consider the two possible weighting schemes.
\begin{itemize}
\item In the unweighted case $d_i=1$ for $i=1,\ldots,n-1$, we obtain from (\ref{eq:Ginoisefirstterm}) that for $i\leq a$:
\begin{equation}\label{eq:Ginoisenoweight}
G_i = \bar{\beta}^2 \frac{i^2 E(n-U)^2 }{n^2} + \frac{i(n-i)}{n}\sigma^2\,.
\end{equation}
While the first term in (\ref{eq:Ginoisenoweight}) is strictly increasing on $[0,a]$, the second term moves the maximum of $G_i$ towards $n/2$. This shows that the maximum of $G_i$ is always at least $a$ when $a\leq n/2$. By symmetry, it is also always smaller or equal to $b$ when $b\geq n/2$. When $n/2 \in [a,b]$, we deduce that for any $\sigma^2>0$, $\hat{u}\in[a,b]$. Otherwise, let us suppose without lack of generality that $n/2 < a \leq b$. Then, $G_i$ being quadratic on $[0,a]$ and equal to $0$ at $0$, the maximum of $G_i$ will not occur before $a$ if and only if $G_{a-1}< G_a$. A computation similar to the one in the proof of Theorem \ref{bigtheorem} shows that
$$
G_a - G_{a-1} = 2  \br{\tilde{\sigma}_m^2 - \sigma^2} \br{\alpha_m - \frac{1}{2} + \frac{1}{2n}}\,,
$$
where
$$
\tilde{\sigma}_m^2 = n \bar{\beta}^2 \frac{E(1-\alpha)^2(\alpha_m - \frac{1}{2n})}{\alpha_m - \frac{1}{2} - \frac{1}{2n}}\,.
$$
This shows that $G_a > G_{a-1}$ if and only if $\sigma^2 < \tilde{\sigma}_m^2$. Since $b>n/2$, we also know that $\hat{u} \leq b$, i.e., $\hat{u} \in [a,b]$ in that case. The case $1 \leq a \leq b < n/2$ can be treated similarly. To conclude the proof it suffices to observe that 
$$
E(1-\alpha)^2 = \br{1- E\alpha}^2 + \text{var}(\alpha)\,.
$$
\item In the weighted case $d_i = \sqrt{\frac{n}{i(n-i)}}$ for $i=1,\ldots,n-1$, we obtain from (\ref{eq:Ginoise}) and (\ref{eq:Ginoisefirstterm}) that for $i\leq a$:
\begin{equation}\label{eq:Ginoiseweight}
G_i = \bar{\beta}^2 \frac{i }{n-i}E\sqb{\frac{n-U}{U}} + \sigma^2\,.
\end{equation}
This is always an increasing function of $i$ on $[1,a]$, showing that the maximum of $G_i$ can not be strictly smaller than $a$. By symmetry, it can also never be larger than $b$, from which we conclude that it is always between $a$ and $b$, i.e., in the support of $P_U$.
\end{itemize}
\qed

\bibliographystyle{unsrt}

\end{document}